%% file: arxiv.tex
\documentclass[manuscript,screen]{acmart}

\usepackage{xcolor}

\usepackage{amsmath,amssymb,amsfonts}
\usepackage{algorithmic}
\usepackage{graphicx}
\usepackage{subcaption}
\usepackage{textcomp}
\usepackage{comment}
\usepackage{hyperref}
\usepackage{bm}
\usepackage{tcolorbox}{\small}

\usepackage{lipsum}
\usepackage{multirow}
\usepackage[justification=centering]{caption}
\usepackage{caption}
\usepackage{url}
\usepackage{listings}
\usepackage{colortbl}
\usepackage{enumitem}
\usepackage{soul}
\usepackage{listings}
\usepackage{makecell}
\usepackage{float}
\hypersetup{hidelinks}

\begin{document}

\title{Uncovering Weaknesses in Neural Code Generation}

\author{Xiaoli Lian}
\email{lianxiaoli@buaa.edu.cn}
\affiliation{%
  \institution{Beihang University}
  \city{Beijing}
  \country{China}
}

\author{Shuaisong Wang}
\affiliation{%
  \institution{Beihang University}
  \city{Beijing}
  \country{China}}
\email{littletree@buaa.edu.cn}

\author{Jieping Ma}
\affiliation{%
  \institution{Beihang University}
  \city{Beijing}
  \country{China}
}

\author{Fang Liu}
\affiliation{%
 \institution{Beihang University}
 \city{Beijing}
 \country{China}}

\author{Xin Tan}
\affiliation{%
 \institution{Beihang University}
 \city{Beijing}
 \country{China}}

\author{Li Zhang}
\affiliation{%
  \institution{Beihang University}
 \city{Beijing}
 \country{China}}
\email{lily@buaa.edu.cn}

\author{Lin Shi}
\affiliation{%
  \institution{Beihang University}
 \city{Beijing}
 \country{China}}
\email{shilin@buaa.edu.cn}

\author{Guiyun Gao}
\affiliation{%
  \institution{Harbin Institute of Technology}
  \city{Shen Zhen}
  \country{China}}
\email{gaocuiyun@hit.edu.cn}

\renewcommand{\shortauthors}{Xiaoli Lian et al.}

\begin{abstract}
Code generation, the task of producing source code from prompts, has seen significant advancements with the advent of pre-trained large language models (PLMs). Despite these achievements, there lacks a comprehensive taxonomy of weaknesses about the benchmark and the generated code, which risks the community's focus on known issues at the cost of under-explored areas.
Our systematic study aims to fill this gap by evaluating five state-of-the-art PLMs—three larger models (CodeGen2.5 (7B), CodeGeeX2 (6B), GPT-4-Turbo) and two smaller ones (UnixCoder (110M) and CodeT5-base (220M))—across three popular datasets (CoNaLa, HumanEval+, and DS-1000). We assess the quality of generated code using match-based and execution-based metrics, then conduct thematic analysis to develop a taxonomy of nine types of weaknesses.
We dissected weakness distributions and uncover three findings: (1) In the CoNaLa dataset, inaccurate prompts are a notable problem, causing all large models to fail in 26.84\% of cases, with even higher failure rates of 40\% for smaller models; (2) Missing pivotal semantics is a pervasive issue across benchmarks, with one or more large models omitting key semantics in 65.78\% of CoNaLa, 66.09\% in HumanEval+ and 80.51\% in DS-1000; (3) All models struggle with proper API usage, a challenge amplified by vague or complex prompts.
\end{abstract}





\maketitle

\input{sections/Introduction}

\input{sections/ResearchSubjects}

\input{sections/ImperfectCaseSelection}

\input{sections/WeaknessTaxonomy}

\input{sections/Distributions}
\input{sections/Discussion}
\input{sections/RelatedWork}
\input{sections/Conclusion}

 \section*{Acknowledgments}
 We extend our gratitude to the additional participants who contributed to data annotation beyond the authors. Our thanks go to Feng Luo, Zixuan Zhai, and Jiehua Li from Harbin Institute of Technology, as well as Jiale Xu, ChengHu, Ziyan Zhao, Donghao Yang, Tiansheng Lin, and Shaoming Zhu, from Beihang University.
 Funding for this work has been provided by the National Science Foundation of China Grant No.62102014 and supported by State Key Laboratory of Software Development Environment No.SKLSDE-2023ZX-10.
\bibliographystyle{ACM-Reference-Format}
\bibliography{arxiv}

\end{document}

%% file: sections/Introduction.tex
\section{Introduction}
\label{sec:intro}

Code generation refers to the task of automatically generating source code in a specific programming language for a given prompt. Automatic code generation has emerged as a pivotal technology in the field of software development, with the potential to reduce errors associated with manual coding and drastically improve productivity. On May 3, 2023, during the Web Summit, the GitHub CEO Thomas Dohmke demonstrated the potential of AI-assisted programming by implementing a snake game within just 15 minutes \footnote{\url{https://rio.websummit.com/sessions/wsr23/0287e463-adf7-412b-8cf1-651de1a6128d/how-to-build-an-app-in-18-minutes/}}. He achieved this by just asking for and copying code generated by GitHub Copilot X, an AI-powered code generation tool. He claimed that each developer can increase their productivity by 10 times with AI. Besides, many companies have launched their own AI-powered coding assistants, such as Microsoft's Visual Studio IntelliCode \cite{Intellicode}, Github Copilot X \cite{CopilotX}, Facebook's Aroma \cite{Aroma} and so on.

In recent years, a series of impressive models have been proposed, such as sequence-based \cite{dong-lapata-2016-language,10.5555/3295222.3295349}, tree-based \cite{yin-neubig-2018-tranx}\cite{chen2020treegen}, and pre-trained code generation models \cite{feng-etal-2020-codebert}\cite{nijkamp2022codegen}\cite{wang2021codet5}\cite{li2023skcoder}, and they are evolving at an unprecedented rate. State-of-the-art (SOTA) code generation models, particularly those based on pre-trained large language models (PLMs), have demonstrated promising performance in generating source code for a variety of programming languages and tasks. However, despite their successes, these models still fail or perform poorly for some cases \cite{liu2020deep}, and understanding their limitations is crucial for driving further advancements.

Recent studies have illuminated various challenges associated with PLMs from multiple angles. Shin et al. \cite{10.1145/3630009} presented a thorough evaluation of six neural code generation models, scrutinizing their performance across four benchmark datasets, and offered a detailed look at model competencies in terms of accuracy, syntax and semantic correctness, as well as the practical utility of the code produced. 
Antonio et al. \cite{mastropaolo2023robustness} assessed GitHub Copilot's robustness, noting that varying yet semantically similar natural language descriptions can prompt different function recommendations. Adding to this, Zan et al. \cite{zan-etal-2023-large} identified three pivotal elements for successful large language models (LLMs) in natural language to code (NL2Code) tasks: extensive model size, high-quality data, and expert fine-tuning. Furthermore, Song et al. \cite{wang2024largelanguagemodelsfail} dissected code generation errors across three LLMs within the HumanEval dataset, categorizing mistakes into specific semantic and syntactic issues, such as incorrect function arguments. One similar job is done by Tambon et al. \cite{tambon2024bugs} on another benchmark CoderEval \cite{10548523}.

While these contributions spotlight key problems like constant value inaccuracies and erroneous function parameters, there is a discernible gap: a comprehensive classification of foundational weaknesses, from the views of both the benchmarks and the generated code, has not been extensively explored.  This oversight
might inadvertently lead the research community to fixate on well-documented challenges, thereby neglecting a host of significant but under-researched issues that merit further exploration.




In this study, we aim to uncover the weaknesses in automatic code generation by conducting an in-depth analysis of the problematic codes produced by five advanced PLMs across three widely-used datasets. In particular, we first collected five SOTA PLMs, including UnixCoder \cite{guo2022unixcoder}, CodeT5-base \cite{wang2021codet5}, CodeGen2.5\cite{nijkamp2022codegen}, CodeGeeX2\cite{CodeGeeX}, and GPT-4\cite{openai2024gpt4}, and then evaluated the generated code quality in three public datasets, including CoNaLa \cite{yin2018mining}, HumanEval+ \cite{chen2021evaluating} and DS-1000 \cite{lai2022ds}. Through thematic analysis on the problematic cases, we identified nine weakness types from the perspectives of both benchmark and PLMs. 
Furthermore, we uncovered how these weaknesses are distributed among the pairs of prompts and the source code generated by different PLMs. Through this examination, we aim to underscore the predominant weakness types unique to distinct benchmarks as well as the frailties specific to various PLMs. 

 \textbf{\emph{Observations.}} Our comprehensive analysis reveals several key observations pertaining to both the benchmarks and the generated code:
 
\textbf{(1) Benchmark Insights:} Within the CoNaLa dataset, the most significant weaknesses identified include single answer biases and inaccurate prompt descriptions. The larger models demonstrate varying and notable degrees of susceptibility to single answer biases with 43.66\%, 32.15\%, and 23.01\% of the evaluated samples. Inaccurate prompts also pose a significant challenge; they cause all three larger models to generate incorrect code in more than 27\% of instances, while smaller models experience an even higher error rate of 40\%. Conversely, within the HumanEval+ and DS-1000 datasets, it is the smaller models that are particularly troubled by complex prompts, with error rates reaching 43.75\% and 27\% respectively in shared problem areas.

\textbf{(2) Generated Code Insights:} Across PLMs, a common deficiency is observed in generating source code that fails to encompass all vital semantic elements from the prompts. This trend indicates that even with sophisticated NLP skills, the larger models tend to miss critical details, especially within extensive descriptions. Additionally, it is noted that larger models may generate more verbose code than necessary, particularly when faced with ambiguous or imprecise prompts.

\textbf{\emph{Suggestions.}} Based on these insights, we propose four suggestions for future research:
\textbf{(1) Prompt refinement:} Reformulate prompts to enhance their clarity and conciseness, thereby reducing inaccuracies and complexities.
\textbf{(2) Sophisticated prompt engineering:} Develop advanced techniques for prompt engineering to ensure PLMs not only grasp the overall meaning but also recognize and retain every crucial detail in the prompts, effectively preventing the omission of essential semantics in the output code.
\textbf{(3) API usage proficiency:} Improve PLMs' proficiency in API usage, especially when dealing with complex prompts.
\textbf{(4) Advanced stop criteria of generation:} Increase the precision of code generation and refine the criteria for stopping, aiming to eliminate unnecessary verbosity (`gold plating') and code duplication.

This work builds upon our preliminary findings presented as a poster at the 46th International Conference on Software Engineering (ICSE 2024), where we introduced the taxonomy of weaknesses without an in-depth exploration. In our current exposition, we provide detailed annotations, exemplify each weakness type, and analyze the distribution of these weaknesses through both individual and collective results of different models. Further, we explore the effects of manually curating inaccurate prompts within the CoNaLa dataset and illustrate how such refinements can enhance code evaluation outcomes. Our preliminary findings support the notion that prompt reformulation can significantly improve the quality of generated source code. Finally, we address the current limitations and highlight new opportunities in the field of neural code generation.

\textbf{Contributions.} The major contributions of this paper are summarized as follows: 
    
\begin{itemize}[leftmargin = 0.4cm, itemindent = 0.1cm]
    \item We reveal a taxonomy consisting of nine unique weaknesses that permeate across five SOTA PLMs and three widely-used benchmark datasets, with an analysis framed by both the benchmarks and the capabilities of the neural models.

    \item Our exploratory study brings to light the adverse effects of imprecision and complexity in prompts on the code generation process, underpinned by quantitative distribution data. We advocate for sustained attention to accurate API usage in the generated code, especially when dealing with complex prompts. Additionally, our analysis identifies three commonly overlooked weakness types: the omission of vital semantics, the introduction of unnecessary complexity, and the incidence of superfluous duplication in the code.
        
    \item Drawing from our findings and existing literature, we articulate four areas where limitations intersect with opportunities in the realm of neural code generation, setting the stage for future advancements in this field.
\end{itemize}

 Our findings not only serve as a foundation for further research in this area but also offer practical guidance for the development and application of more robust and reliable code generation models. The generated code and our annotations are accessible via \url{https://figshare.com/s/97fc12f62929b9b88d85}.

%% file: sections/ResearchSubjects.tex
\section{Research Subjects}
\label{sec:modelReproduce}

\subsection{Models}

 Generally, code-generation models can be grouped into three categories: sequence-based, tree-based and pre-trained models \cite{li2023skcoder}. And the outperformance of pre-trained models, especially the recently proposed ones, have been shown by multiple research \cite{li2023skcoder, liu2020deep}. Therefore, we selected five advanced code-PLMs, including  UnixCoder \cite{guo2022unixcoder}, CodeT5-base \cite{wang2021codet5}, CodeGen2.5\cite{nijkamp2022codegen}, CodeGeeX2\cite{CodeGeeX} and GPT-4\cite{openai2024gpt4}. 
 \begin{itemize}[leftmargin = 0.2cm, itemindent = 0.1cm]
     \item \textbf{UniXCoder} (110M) \cite{guo2022unixcoder} is a multi-layer Transformer model, and it uses AST and code comment to enhance code generation. The advantage has been shown on five code-related tasks over nine datasets.
     \item \textbf{CodeT5-base} (220M) \cite{wang2021codet5} is an encoder-decoder Transformer model, in which a novel identifier-aware pre-training task is proposed to enable better identifier recovery. It has achieved SOTA results on code understanding and generation over one popular benchmark.
     \item \textbf{CodeGen2.5} (7B) \cite{nijkamp2023codegen2} is a multi-turn program synthesis model, which progressively accepts specifications in natural language and generates the subprograms. It outperfoms CodeGen1.0-16B-mono and CodeT5-16B-mono models with only 7B on HumanEval benchmark.
     
     \item \textbf{CodeGeeX2} (6B) represents the second generation of CodeGeeX \cite{CodeGeeX}, a multilingual model designed specifically for code generation tasks. Despite housing only 6 billion parameters, this advanced model exhibits superior performance in code generation when tested on two popular benchmarks - HumanEval \cite{chen2021evaluating} and DS-1000 \cite{lai2022ds}. Remarkably, it outperforms several larger-scale, sophisticated models, including but not limited to OpenAI Codex-12B \cite{chen2021evaluating}, LLaMA2-70B \cite{touvron2023llama}, and StarCoder-15B \cite{li2023starcoder}.
     
     \item \textbf{GPT-4-0613}(GPT-4) \footnote{https://openai.com/gpt-4}\cite{openai2024gpt4} is  the state-of-the-art fourth iteration of OpenAI's Generative Pre-trained Transformer series. It boasts an expanded parameter count, granting it an advanced capability to comprehend and process intricate instructions, thereby establishing itself as a formidable baseline for code generation applications.
 \end{itemize}

 To ensure the correctness of the first four models' implementation, we directly downloaded and used the source code from their official websites. For GPT-4, we interact with it via a human-model dialogue interface, enabling an intuitive engagement and seamless exchange of information.
 


\subsection{Benchmark Datasets}

For a comprehensive evaluation, we choose two types of datasets based on their evaluation methods: match-based evaluation (i.e., CoNaLa) and unit-test-based (i.e., HumanEval+ and DS1000). The relevant information is presented in Table \ref{tab:dataset}. These datasets consist of prompt-code pairs, primarily focusing on method-level granularity for the code.

\begin{itemize}[leftmargin = 0.2cm, itemindent = 0.1cm]
    \item \textbf{CoNaLa} \cite{yin2018mining} \cite{Conala} is a collection of 2,879 pairs of how-to question and Python code fragments mined from QA forums (e.g, Stack Overflow \cite{SO}). The original team also conducted a curated processing to improve the quality of CoNaLa by incorporating variable names and function arguments into the prompts. Given that our task is to build a taxonomy of the weakness of SOTA models and high-quality datasets, we select the rewritten version.
    
    \item \textbf{HumanEval+}  is an enhanced variant of the broadly utilized benchmark, HumanEval \cite{chen2021evaluating}. This benchmark comprises 164 programming problems and Python functions, accompanied by unit tests. In contrast to its predecessor, HumanEval+ expands the test-cases nearly 81-fold. This substantial enhancement allows for a more comprehensive evaluation of the generated code's functional correctness. Furthermore, it significantly mitigates potential biases that might arise from single-answer evaluations.
    
    \item  \textbf{DS-1000} \cite{lai2022ds} is a code generation benchmark with 1000 problems originating from 451 Stack Overflow problems. The most advantage is that it contains prompts spanning seven widely-used Python data science libraries: NumPy, Pandas, Tensorflow, PyTorch, SciPy, Scikit-learn, and Matplotlib.

\end{itemize}


\begin{table}[!htbp]
\caption{Benchmark Datasets Information}
\label{tab:dataset}
\footnotesize
\centering
\begin{tabular}{|c|c|c|c|}
\hline
\textbf{Name} & \textbf{CoNaLa}  & \textbf{HumanEval+}  & \textbf{DS-1000} \\ \hline
Year &2018&2021& 2022 \\ \hline
Source &\cite{yin2018mining}&\cite{chen2021evaluating}&\cite{lai2022ds} \\ \hline
\#Prompts &2,879&164&1000 \\ \hline
\#Avg.Test Case &0 & 7.7 & 1.6  \\ \hline
Average tokens in prompts & 10.06 &67.72&137.01\\ \hline
Average tokens of programs & 4.33 &24.38&13.87 \\ \hline
\end{tabular}
\end{table}

%% file: sections/ImperfectCaseSelection.tex
\section{Case Selection and Annotation}
\label{sec:badCaseSelection}


To construct the weakness taxonomy, we initially assess the generated code to identify problematic instances. Subsequently, we conduct a thematic analysis of the prompts, reference answers, and generated code from these selected models.

\subsection{Code Generation Evaluation}
\label{subsec:rep}
We evaluate the generated source code to select the problematic cases, i.e., the generated code can't resolve the given prompts.

\noindent \textbf{Metrics} We select three popular metrics, including two match-based metrics and one (unit-test) execution-based metric, to evaluate the quality of generated source code. 
 \begin{itemize}[leftmargin = 0.2cm, itemindent = 0.1cm]
  \item \textbf{Exact match (EM)} is the ratio of the generated code that is exactly same with the reference answer, i.e., with the same token sequence. 
     \item \textbf{CodeBLEU} \cite{ren2020codebleu} is also a match-based metric. It not only considers the n-gram match (i.e., BLEU), but also the syntactic match via abstract syntax trees (ASTs) and code semantic match via data-flow. We integrate BLEU-4 \cite{papineni2002bleu} in this present work, which measures the 4-gram similarity between the generated code and the ground-truth. It is widely employed by existing approaches to evaluate the quality of code generation \cite{li2023skcoder, liu2020deep}.
     \item  \textbf{Pass@K} calculates the ratio of resolved prompts. For one prompt, \emph{k} source code samples would be generated, and the prompt is regarded as resolved if any sample can pass the all prompt-related unit tests. In this work, we use \emph{Pass@1} because users of code-PLMs usually only need a helpful one in real practical scenario and it also has been widely used in the existing studies \cite{li2023skcoder, zan2022cert}.
 \end{itemize}
The ranges of all the above metrics are [0,1], with higher values being better. EM and pass@1 were used to identify the fully correct source code for certain prompts. When the answers are not satisfied 100\%, CodeBLEU was used to evaluate the quality of generated source code, comparing with the given answers. 
To ensure the correct implementation, for these three evaluation metrics, we utilized a popular online implementation of the comprehensive CodeBLEU available on GitHub \footnote{\url{https://github.com/mahimanzum/FixEval}} and Python libraries for the other two metrics. 

\vspace{0.4em}
\noindent \textbf{Setup.} Given the scale of the parameters and the similarity between the pre-training datasets of the models to our selected benchmarks, we decided to fine-tune the two smaller models, UnixCoder (110M) and CodeT5-base (220M), across all three datasets. We directly used CoNaLa's train/valid/test divisions for the corresponding tasks, comprising 500 test prompt-code pairs. For the other two benchmarks that lacked these partitions, we employed an 8:1:1 ratio. In other words, 16 samples of HumanEval+, and 100 of DS-1000 were used for testing the two smaller models.

The three larger models are theoretically capable of handling all samples directly due to the reduced necessity for fine-tuning. However, the task of annotating five sets of responses for each prompt in the comprehensive datasets—encompassing 2879 samples from CoNaLa, 164 from HumanEval+, and 1000 from DS-1000—presents a prohibitive time commitment. Consequently, we selectively annotate only a subset of samples for thorough examination with these larger models.

For a 95\% confidence level and a 5\% margin of error, we determined the necessary sample sizes to be 339 for CoNaLa, 115 for HumanEval+, and 277 for DS-1000. We then randomly drew corresponding samples from these benchmarks, executed the three larger models, identified problematic cases based on our metrics, and carried out annotations accordingly.

Notably, the involving three larger models, i.e., CodeGen2.5, CodeGeeX2, and GPT-4 often produce supplementary comments and functions besides the answer code. For a more accurate evaluation, we applied the respective common truncation strategies to the outputs from CodeGen2.5 and CodeGeeX2, as detailed in \cite{nijkamp2023codegen2, bavarian2022efficienttraininglanguagemodels}. For GPT-4's output, since the human-machine interaction process is conducted, we manually filtered out any extraneous content for a cleaner analysis.

\noindent \textbf{Results.} The performance of five PLMs over the three datasets is shown in Table \ref{tab:results}. We can make two observations. 

An examination of each benchmark's outcomes reveals that on CoNaLa, where the two smaller models underwent additional fine-tuning, they outperform the three larger models when evaluated using only match-based metrics. Within HumanEval+, CodeGeeX2 secures the top position in match-based metrics, while GPT-4 excels with the highest \emph{pass@1} rate. Meanwhile, in DS-1000, GPT-4 achieves the best \emph{EM} and \emph{pass@1} rates, and UnixCoder earns the highest score in \emph{CodeBLEU} metric. These findings suggest that with a suitably sized dataset for fine-tuning (2,379 samples for CoNaLa, and 800 for DS-1000), smaller models can surpass their larger counterparts in effectiveness.

When assessing the performance of various PLMs across different benchmarks, it's apparent that GPT-4, exhibits superior \emph{pass@1} rates in both HumanEval+ and DS-1000. Nevertheless, it appears less effective in match-based metrics across all three evaluated benchmarks. For instance, in HumanEval+, while GPT-4's \emph{EM} stands at 0, its \emph{pass@1} rate is notably high at 66.96\%. This highlights GPT-4's creative capabilities, reflecting its inclination to generate diverse yet accurate answers.
In DS-1000, although GPT-4's \emph{CodeBLEU} score falls short compared to UnixCoder, its \emph{pass@1} rate substantially exceeds that of UnixCoder. Besides, CodeGeeX2 shows better performance than CoderGen2.5, even with a smaller parameter scale, in all of the three benchmarks.



\begin{table}[!htbp]
	\caption{Performance of PLMs over the Three Datasets}
	\footnotesize
	\label{tab:results}
	\centering
	\renewcommand\arraystretch{1.3} 
	\begin{tabular}{|c|c|p{0.5cm}|p{1.3cm}|c|}\hline
		\textbf{Benchmark} & \textbf{Models} &  \textbf{EM} & \textbf{CodeBLEU}  & \textbf{Pass@1}\\ \hline 
		\multirow{4}{*}{CoNaLa} &UnixCoder &6.80&30.50
		&- \\ \cline{2-5}
		& CodeT5 & \textbf{7.40} & \textbf{31.10}  
		&-\\ \cline{2-5}
		&CodeGen2.5 &0 & 26.70  & - \\ \cline{2-5}
		&CodeGeeX2 &0.29 &28.83 &- \\ \cline{2-5}
		&GPT-4 &0 & 22.22 & - \\ \hline \hline
		\multirow{4}{*}{HumanEval+} &UnixCoder &0&18.45&0 \\ \cline{2-5}
		& CodeT5 &0&18.03 &0\\ \cline{2-5}
		&CodeGen2.5 &0.87&25.86&20.87 \\ \cline{2-5}
		&CodeGeeX2 &\textbf{6.09} &\textbf{30.32} &28.70 \\ \cline{2-5}
		&GPT-4 &0 &28.30 &\textbf{66.96} \\ \hline \hline
		\multirow{4}{*}{DS-1000} &UnixCoder &5.00&\textbf{51.59}&11.00 \\ \cline{2-5}
		& CodeT5 &5.00&37.82 &11.00\\ \cline{2-5}
		&CodeGen2.5 & 0.72 & 20.55 & 5.78 \\ \cline{2-5}
		&CodeGeeX2 &5.78 &24.01 &22.38 \\ \cline{2-5}
		&GPT-4 &\textbf{6.14} & 26.69 & \textbf{29.60} \\ \hline \hline
	\end{tabular}
\end{table}


\subsection{Thematic Analysis of Problematic Cases}

We employ a thematic analysis process \cite{braun2022conceptual} to build the weakness taxonomy. We five authors form the initial group. And we recruited additional participants through fourth-year undergraduate and master CS courses at Beihang University and Harbin Institute of Technology. We conducted brief interviews and selected students with more than two years of Python development experience. In the end, we chose nine more students to join our team. These students, alongside five authors, formed an annotation group with 14 participants, with three tenured CS assistant professors and 11 CS PhD/master/undergraduate students. 

Our analysis includes three rounds.
We began with a preliminary study aimed at identifying an initial range of weakness categories to serve as a foundation for our thematic analysis. To do this, we first organized the code generated for CoNaLa by UnixCoder and CodeT5 based on their \emph{CodeBLEU} scores, provided that their \emph{EM} score was not 1. Next, we isolated the bottom 10\% of cases for each PLM, seeking commonalities by pinpointing prompts that were consistently found in the poorest performing subset across all PLMs. Through this process, we identified 33 cases. 

The initial analysis of these 33 cases was performed independently by the first two authors, who delved deeply into each instance. They compared the original prompts, reference code, and the responses produced by the five PLMs, from which they derived an initial taxonomy of weaknesses. Following this, the first three authors convened to thoroughly discuss and refine these categories, ultimately agreeing upon an enhanced framework for classifying the identified weaknesses. In this phase, a category with six weakness types was built.

We then identified all problematic testing samples for each PLM to be annotated. Specifically, we selected instances in CoNaLa with an \emph{EM} score of less than 1, and cases in HumanEval+ and DS-1000 where the \emph{pass@1} rate was below 1. For the two smaller models, all testing samples were deemed problematic as none produced perfect source code that either exactly matched the reference code in CoNaLa or successfully passed all given test cases in HumanEval+ and DS-1000. The number of annotated samples for each PLM across the benchmarks is detailed in Table \ref{tab:annotatedSample}. For the two smaller models, the number of samples from the three benchmarks is 476, 16, and 100, respectively.

For the three larger models—CodeGen2.5, CodeGeeX2, and GPT-4—the total number of testing cases for CoNaLa is 339, encompassing all such cases. In HumanEval+ and DS-1000, where test cases accompany the prompts, the number of problematic samples is fewer than the total number of testing cases. Notably, of the 115 testing samples in HumanEval+, 91 generated by CodeGen2.5, 82 by CodeGeeX2, and 38 by GPT-4 were problematic and therefore selected for annotation. Out of the 277 testing samples in DS-1000, 261 cases generated by CodeGen2.5 were problematic and chosen for annotation, along with 215 for CodeGeeX2 and 189 for GPT-4.

In total, we annotated 3,125 prompt-code pairs. For each pair, we made certain that annotation was independently performed by two participants. During this process, annotators were required to consult the initial set of weakness categories established by our first three authors, contemplate the underlying rationale, and assign a weakness type to each individual prompt-code pair. This type could either align with the pre-existing categories or represent a newly identified category outside the initial framework. Upon receiving the independent annotations, one of our authors reviewed them for consistency. In cases of discrepancy, we facilitated a joint face-to-face discussion between the two annotators to delve into their rationales and ultimately reach a consensus on the weakness type. This approach of independent annotation followed by collaborative reconciliation aimed to minimize subjectivity in the annotation process.

Subsequently, throughout the latter stages of the annotation process, three additional types of weaknesses were recognized. Consequently, we have developed \emph{a comprehensive taxonomy with nine weakness types} in code automatically generated by five PLMs across three widely-used benchmarks. 

\begin{table*}[!htbp]
\caption{The Distribution of Annotated Sample Sizes Across Each Benchmark for Different Models.}
\label{tab:annotatedSample}
\centering
\renewcommand\arraystretch{1.3}
\begin{tabular}{|p{3cm}|p{2.5cm}|p{1cm}|p{1.5cm}|p{1.1cm}|p{4.2cm}|}
\hline
\multirow{2}{*}{\textbf{Study}} & \multirow{2}{*}{\textbf{Involving Models}} & \multicolumn{3}{c|}{\textbf{Number of Samples}} & \multirow{2}{*}{\textbf{Annotation Procedure}}  \\ 
\cline{3-5}
& & CoNaLa & HumanEval+ & DS-1000 & \\ \hline
Pilot study & UnixCoder, CodeT5 & 33 & - & - & Collaborative annotation by the first three authors \\ \hline
Smaller Model Annotation & UnixCoder, CodeT5 & 476 & 16 &100 & Two separate annotations, then one joint review \\ \hline
\multirow{3}{*}{Large Model Annotation} & CodeGen2.5 & 339 & 91 & 261 & \multirow{3}{*}{\shortstack{Two separate annotations, then \\one joint review}} \\ \cline{2-5}
& CodeGeeX2& 339 &82 &215 &\\ \cline{2-5}
& GPT-4& 339 &38 &189 &\\ \hline 

\end{tabular}
\end{table*}



%% file: sections/WeaknessTaxonomy.tex
\section{Weakness taxonomy}
\label{sec:badCaseAnalysis}

\begin{figure*}[!htbp]
	\centering
	\includegraphics[trim = {1cm 4cm 1.5cm 1cm }, clip,  width=0.95\textwidth]{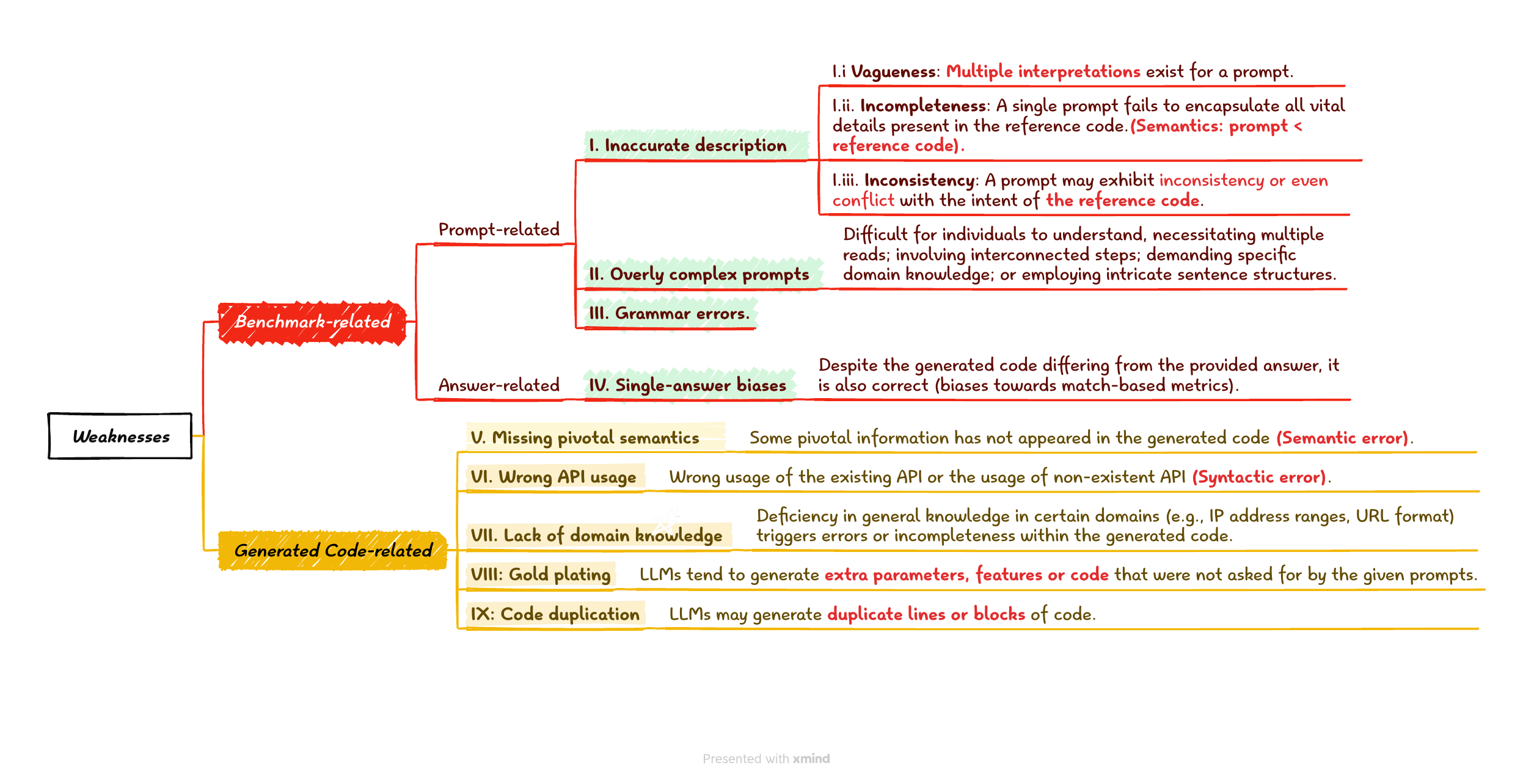}
	\caption{The Overview of the Weakness Taxonomy.}
	\label{fig:taxonomy}
\end{figure*}

We classified the nine identified weakness types into two categories based on their association with either the benchmarks or the generated code, as depicted in Fig. \ref{fig:taxonomy}. Since a benchmark consists of prompt and reference code pairs, we further divided the benchmark-related weaknesses into subcategories pertaining to prompts and reference answers.

It is imperative to underscore that our goal is to identify ``weaknesses'' in the low-quality source code generated by PLMs in response to specific prompts, as well as in comparison with the provided reference code. In essence, our focus lies on the \emph{manifestation} of these weaknesses rather than investigating the ``root causes'' behind the suboptimal code generation. Additionally, it is critical to recognize that a single prompt-code pair may contain several types of weaknesses.


\subsection{Benchmark-related weaknesses}

\subsubsection{Prompt-related weaknesses} 
\label{subsec:promptWeakness}

There are three types of weaknesses associated with the prompts.

\textbf{Type I: Inaccurate description.} If the prompt cannot describe the intent accurately, any PLM is almost impossible to generate the accurate source code. 

We regard the reference source code carries the complete intent, and this kind of weakness is identified by comparing the given prompt and reference source code for the selected problematic cases. 
There are three sub-types.

\begin{figure}[!htbp]
    \centering
    \includegraphics[trim={0 3cm 16.5cm 0},clip, width = 0.45\textwidth]{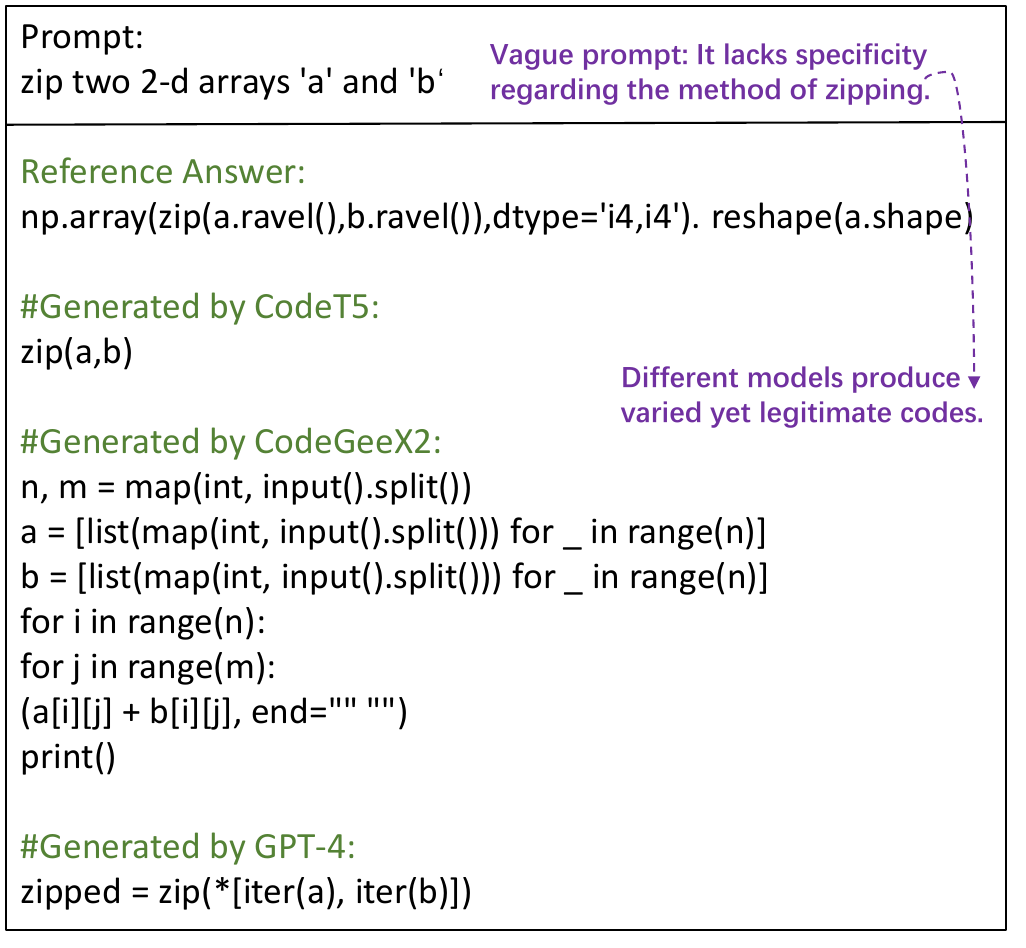}
    \caption{An Example of Diverse Responses Generated by Different Models Due to Vague Prompts.}
    \label{fig:vaguePrompt}
\end{figure}

\emph{I.i. Vagueness.} The inherent vagueness of the prompt introduces a lack of clarity that can give rise to divergent interpretations by different LLMs, resulting in a variety of responses based on each model's unique understanding. In the following example in Fig. \ref{fig:vaguePrompt}, the prompt instructs to zip two 2D arrays `a' and `b'. However, it does not tell the specific zip way, that might lead to confusion or incorrect implementations during code generation. 

Given a = [[2,3], [4,5], [6,7]] and b = [[6,7], [8,9], [10,11], [12,13]], there are several valid methods to zip them. One interpretation is element-wise zipping. It involves first flattening these 2D arrays into 1D arrays, zipping them together, and then reshaping the result to match the original structure of the first array. Following this approach, the reference code yields a zipped result of [[(2,6), (3,7)], [(4,8), (5,9)], [(6,10), (7,11)]].

Another comprehension is row-wise zipping. It allows for an easy zip operation resulting in [[2,3], [4,5], [6,7],[6,7],[8,9]], as implemented by CodeT5 and GPT-4. Additionally, CodeGeeX2 add the elements together, implementing a third ``zip'' operation.




\emph{I.ii. Incompleteness.} Sometimes, the prompt does not encompass all essential details found in the reference code and required \emph{for clear intent}. Generally overlooked elements include
critical parameters (such as a specific string instance), the storage constraint of outcomes, as well as routine operations like ``print''. 

Although this information can be seamlessly integrated into prompts or source code, its omission impedes the execution of precise code implementations. This likely leads to decreased performance across both match-based metrics and execution-based pass@k rates. While addressing such deficiencies might appear straightforward in simpler scenarios—such as within the prompt-code pairs of existing benchmarks—the challenge is significantly magnified in more complex and expansive contexts. In these instances, rapidly detecting buggy code and pinpointing missing elements can become an exceedingly arduous task.

\begin{figure}[!htbp]
    \centering
    \includegraphics[trim={0 3cm 13cm 0},clip, width = 0.5\textwidth]{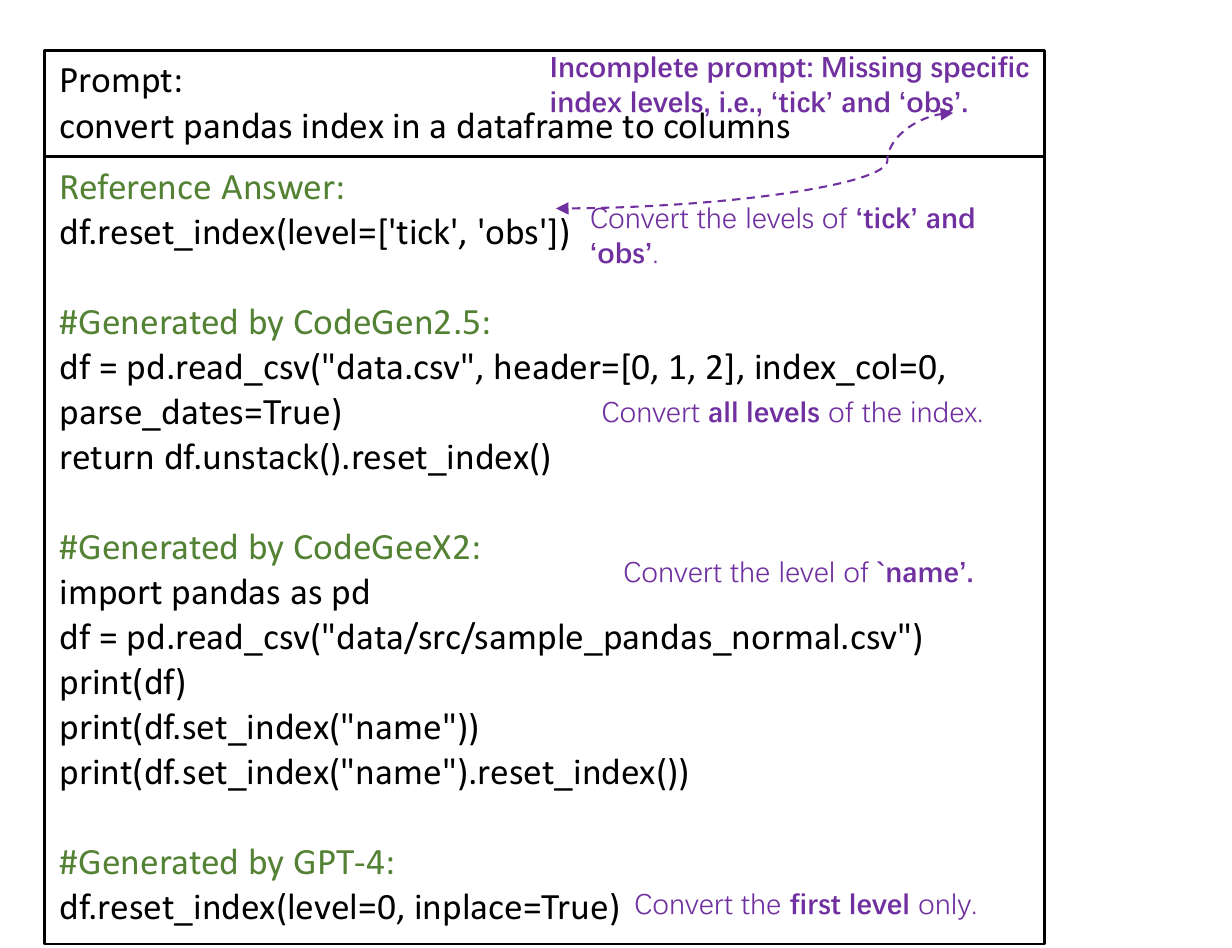}
    \caption{An Example Demonstrating How Incomplete Prompts Result in Varied Outputs Across Different Models.}
    \label{fig:incompletePrompt}
\end{figure}

In the example in Fig. \ref{fig:incompletePrompt}, the prompt does not mention the level names ``tick'' and ``obs'', which leads to two issues:

First, it is unclear whether the DataFrame in question has a MultiIndex or a single-level index, as the prompt does not specify this. Without this information, it is impossible to determine the exact structure of the DataFrame's index.

Second, without explicit level names provided in the prompt, it remains uncertain which specific levels of the index are intended to be converted into columns. This lack of clarity could lead different models to interpret the task differently. For instance:
CodeGen2.5 might interpret the instruction as converting all levels of the index, regardless of their names.
CodeGeeX2 seems to assume that the index has a level named ``name'' and processes the conversion based on that assumption.
GPT-4, on the other hand, interprets the prompt as requiring the conversion of only the first level of the index.

\emph{I.iii. Inconsistency.} At times, prompts may display inconsistencies, or they may even conflict with the reference code provided. This can cause difficulties for PLMs that rely on clear and coherent prompts to accurately generate the intended source code, which are judged by comparing with the reference code.

\begin{figure}[!htbp]
    \centering
    \includegraphics[trim={0 9cm 13cm 0},clip, width = 0.5\textwidth]{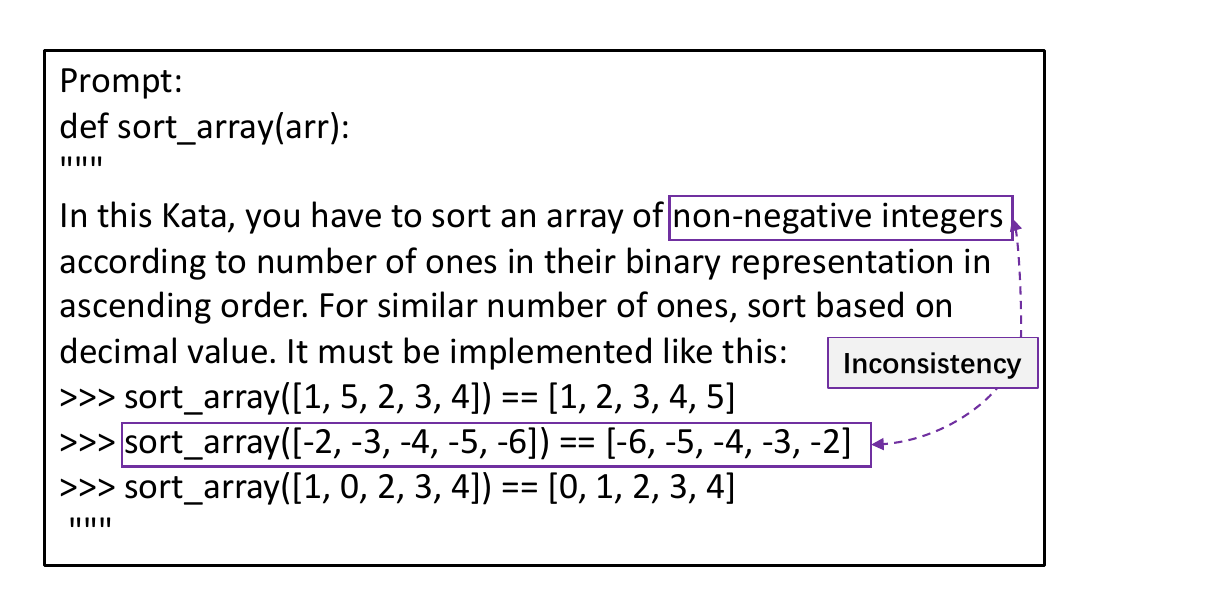}
    \caption{Example of a Prompt with Inconsistency.}
    \label{fig:inconsistentPrompt}
\end{figure}

Consider an example from the HumanEval+ dataset wherein the prompt specifically states that the array to be sorted consists exclusively of non-negative integers, shown in Fig.\ref{fig:inconsistentPrompt}. However, contrary to this specification, the provided second example include negative integers. This discrepancy could lead to confusion for PLMs, which rely on the clarity and accuracy of the prompt to inform their code generation processes. It's essential for the integrity of the prompt and test cases to match so that models can generate appropriate and correct solutions.

\noindent \textbf{Type II: Too complex prompts. }  Intuitively, if prompts are excessively challenging to comprehend, it becomes increasingly difficult for PLMs to generate correspondingly high-quality source code.

During our annotation process, an annotator deems a prompt as overly complex when: 1) repeated reading is necessary to grasp its meaning; or 2) it involves several intertwined steps; or 3) understanding the prompt requires specific domain knowledge; or 4) it has convoluted sentence structures (more subordinate clauses and prepositional phrase). If both annotators agree on the high complexity of a prompt, we label it as ``too complex''. 

\begin{figure}[!htbp]
    \centering
    \includegraphics[trim={0 1cm 16cm 0},clip, width = 0.5\textwidth]{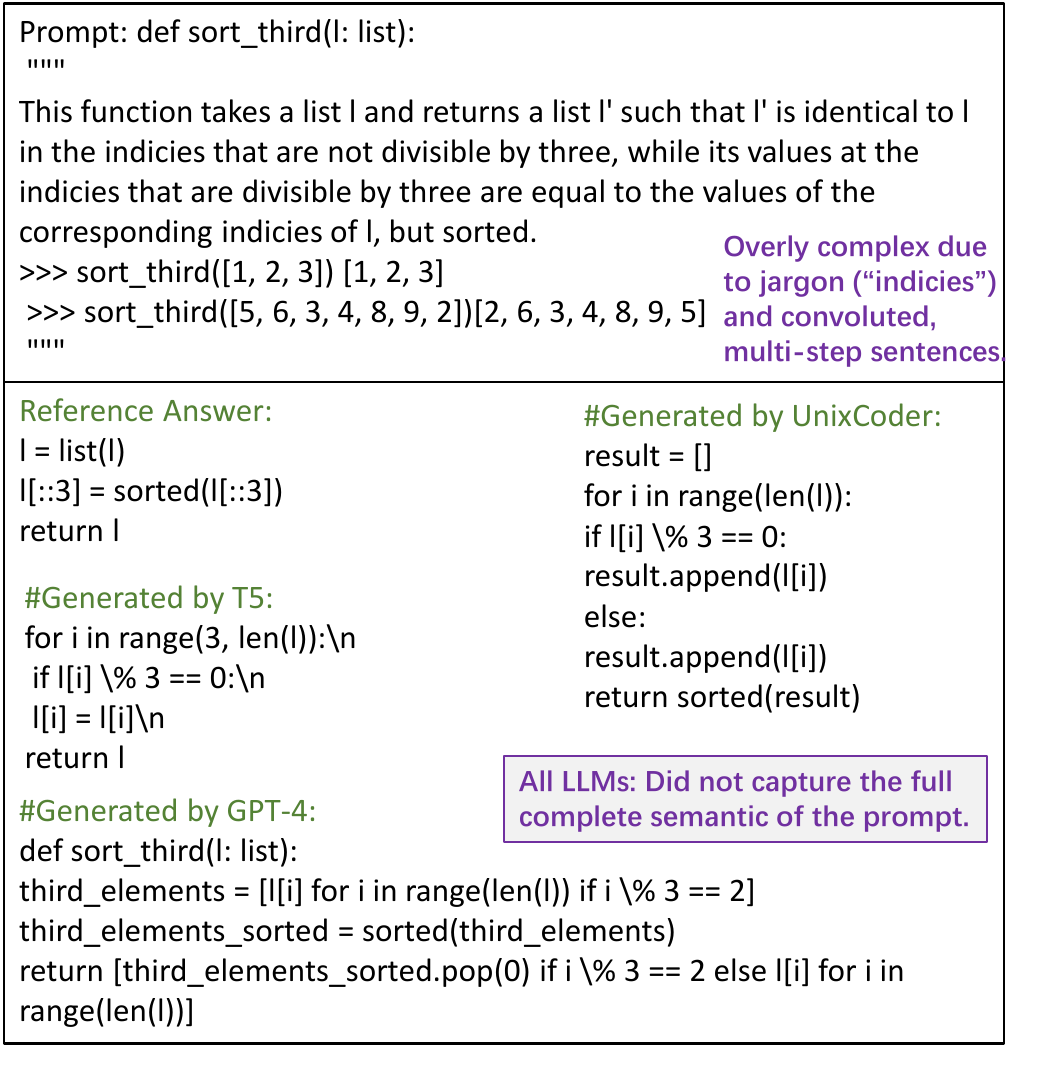}
    \caption{Example of a Overly Complex Prompt.}
    \label{fig:complexPrompt}
\end{figure}

In the provided example in Fig.\ref{fig:complexPrompt}, the prompt incorporates an input (list l), an output (list l'), and two primary steps. Moreover, the second step entails a nested process, specifically sorting. Additionally, to fully comprehend the prompt, annotators must also be well-acquainted with the definition of ``indices''. We list the answers of CodeT5, UnixCoder and GPT-4.

It is evident that there were misunderstandings in the interpretation of the given prompt by CodeT5. It mistakenly replicates only those elements divisible by three. This not only shows a lack of comprehension regarding the meaning of ``indices'' but also overlooks the complexity of the task, including the required two-step process with its nested operations.

UnixCoder also deviated from the task requirements. The first error lies in misconstruing the term ``indices'' to instead operate on the elements themselves. Additionally, UnixCoder failed to accurately identify the sorting criteria, which specifically applies to values at indices divisible by three.

Even GPT-4, with its generally superior interpretive capabilities, was not immune to mistakes. It demonstrated a misunderstanding of the terms ``indices'' and ``locations'', conflating them and thus misinterpreting instructions related to the i-th elements within the list. This confusion led to errors in identifying which elements should be processed according to the prompt's requirements.

These oversights underscore the difficulties that PLMs encounter when attempting to accurately grasp and execute complex prompts, particularly those containing nested procedural steps. The intricate nature of such tasks demands a high level of comprehension capability, which continues to be a significant challenge for PLMs in coding applications.


\noindent \textbf{Type III. Grammar errors.} We find 7 prompts with grammar errors, mainly in CoNaLa. Grammar errors make confusion for automatic prompt comprehension, and then impede the code generation.  




\subsubsection{Answer-related weaknesses}

\begin{figure}[htbp]
    \centering
    \includegraphics[trim={0 1cm 15cm 0},clip, width = 0.5\textwidth]{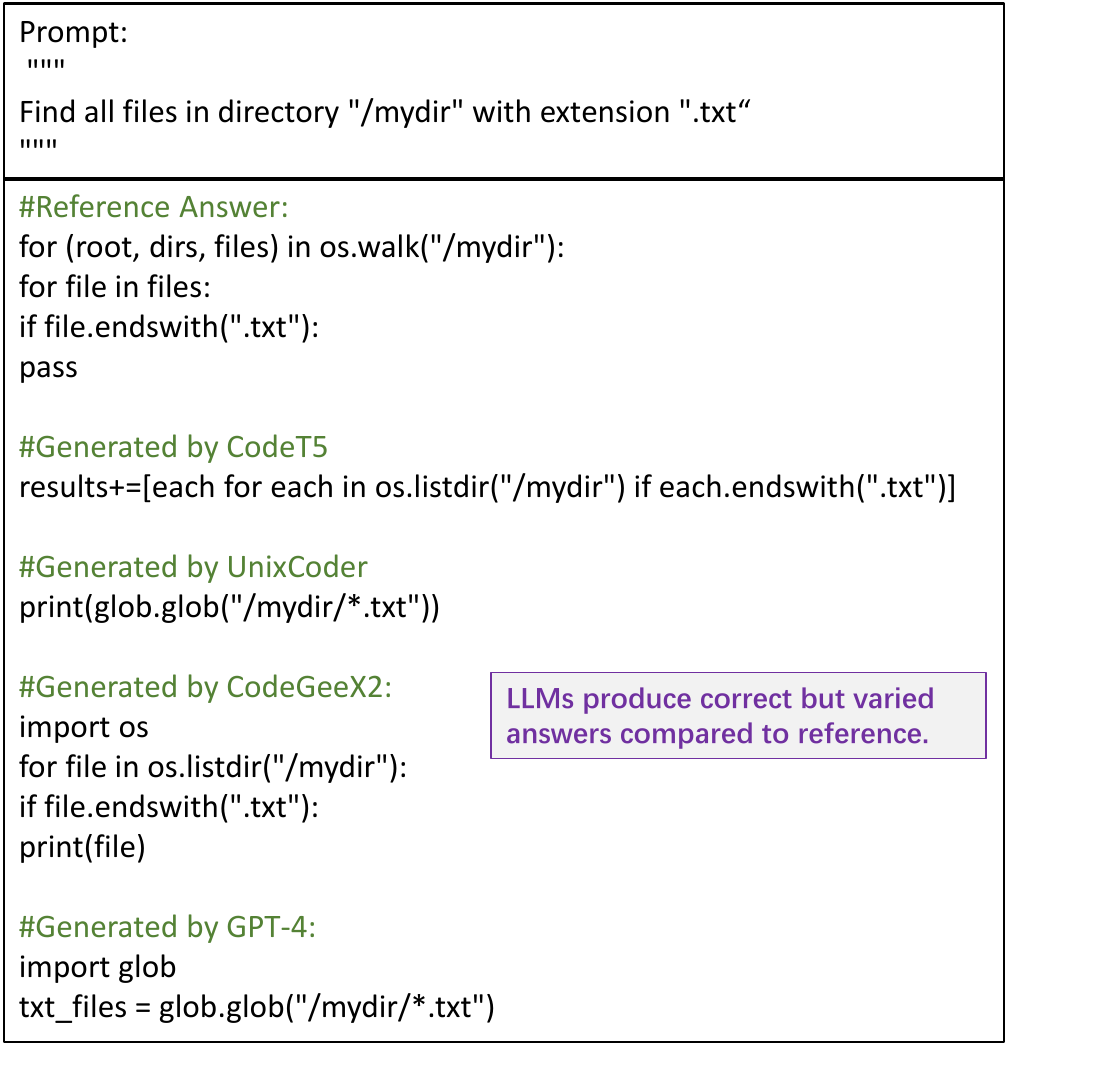}
    \caption{Example Showing How Match-Based Metrics May Misjudge Correct Diverse Responses.}
    \label{fig:singleAnswerPrompt}
\end{figure}

\noindent\textbf{Type IV: Biases from single answer.}  We discovered that 33.70\% problematic cases represent false negatives in CoNaLa, for instance in Fig.\ref{fig:singleAnswerPrompt}. This means that although the generated source code appears different from the golden answer, it is still correct but with a different approach, such as API usage.  

While test case-based measurements like \emph{pass@k} can alleviate this issue, they only work for 
the datasets with test cases. Even in these datasets, when the generated code cannot pass all tests, match-based metrics still hold significance (e.g., suggesting the optimal answer for more effortless modification). In addition, in practical scenarios where test cases are not available (e.g., statement-level code generation), and in widely-used no-test case datasets like CoNaLa and Concode, this remains a concern. 


\subsection{Generated code-related weaknesses}
\label{subsec:modelWeakness}
As shown in Fig. \ref{fig:taxonomy}, we identified five kinds of model-related weaknesses (NO.V - NO.IX), including two LLMs particular weaknesses (NO.VIII and IX).

\noindent \textbf{Type IV. Missing pivotal semantics.} Some pivotal information in the intention prompt has not appeared in the generated code. Therefore, the generated code can not implement the prompt intention. 

During annotation, annotators would juxtapose each generated code against its corresponding prompt. If the generated code failed to incorporate the \emph{central predicate or key constraints} (such as unique formats like \emph{CP 1251 character}), we would consider the associated model to have missed or misunderstood critical semantics. 

\begin{figure}
    \centering
    \includegraphics[trim={0 0cm 16cm 0},clip, width = 0.5\textwidth]{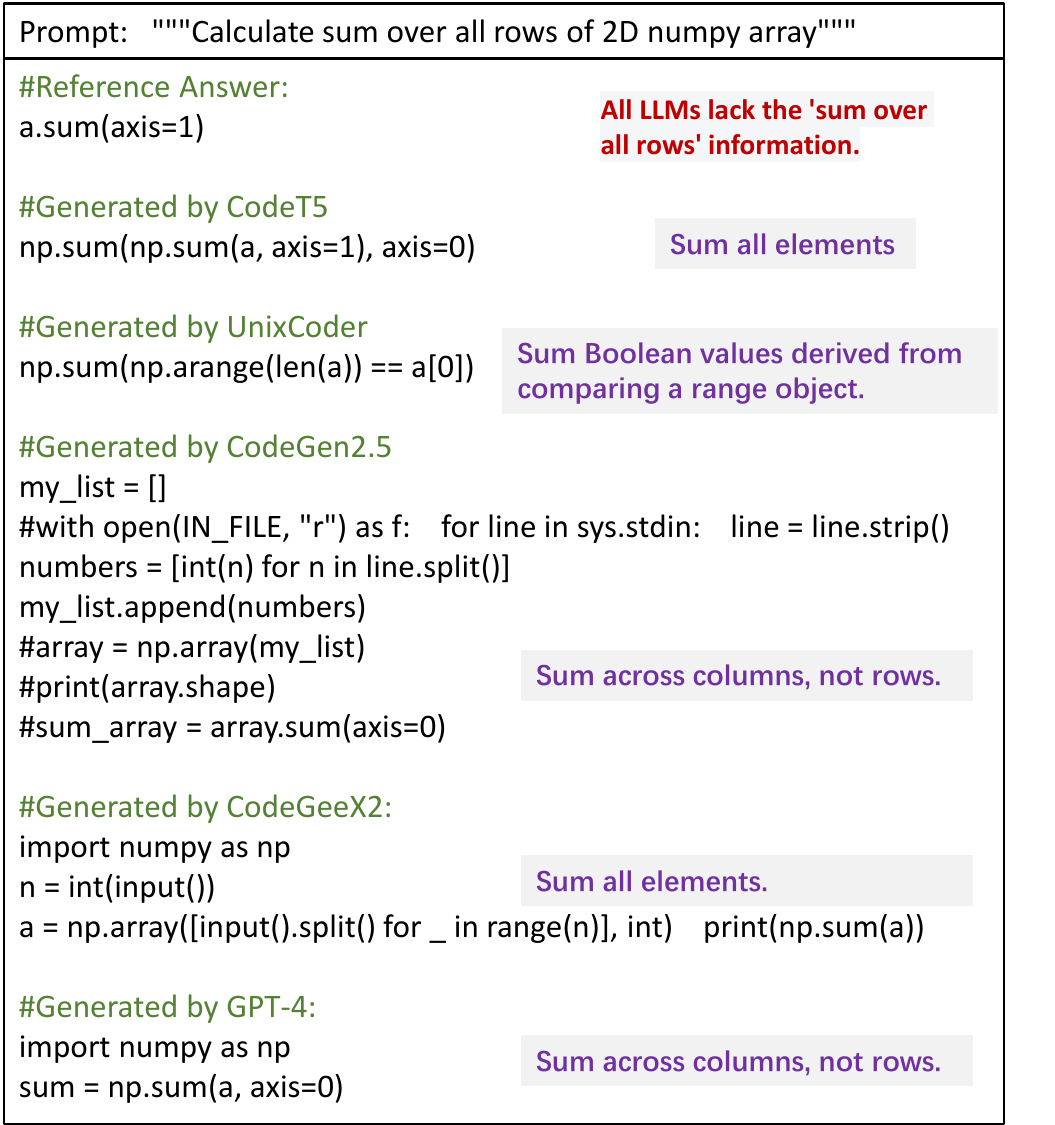}
    \caption{An Example Where the LLMs Overlooks Critical Semantics of the Prompt.}
    \label{fig:missInfoPrompt}
\end{figure}

For instance, consider the following example in Fig.\ref{fig:missInfoPrompt} where all evaluated models—namely, the smaller CodeT5 and UnixCoder, alongside the larger CodeGen2.5, CodeGeeX2, and GPT-4—fail to accurately capture the semantics of ``sum over all rows.'' UnixCoder incorrectly attempts to sum boolean values derived from comparing a range object (whose length matches the number of elements in array a) with the first row of \emph{a}. CodeT5, on the other hand, sums all elements in the array using two instances of np.sum(), rather than computing the row-wise sum as intended. CodeGen2.5 has commented out crucial parts of the code, including the transformation into a numpy array and the sum calculation; however, even if these lines were active, it mistakenly calculates the sum across columns, not rows, in the 2D numpy array. CodeGeeX2 also errs by totaling the elements of the entire array instead of summing each row individually. Similarly, ChatGPT, like CodeGen2.5, incorrectly computes the columnar sum rather than the row-wise sum.

\noindent \textbf{Type V. Wrong API usage.} This weakness pertains to the incorrect use of an existing API (i.e., correct API names but incorrect parameters) or the use of non-existent APIs, reflecting a type of \emph{syntactic error}. Therefore, instances where incorrect APIs are employed correctly but cannot solve the problem indicated in the prompt are not classified under this category. 

\begin{figure}[!htbp]
    \centering
    \includegraphics[trim={0 13cm 16cm 0},clip, width = 0.5\textwidth]{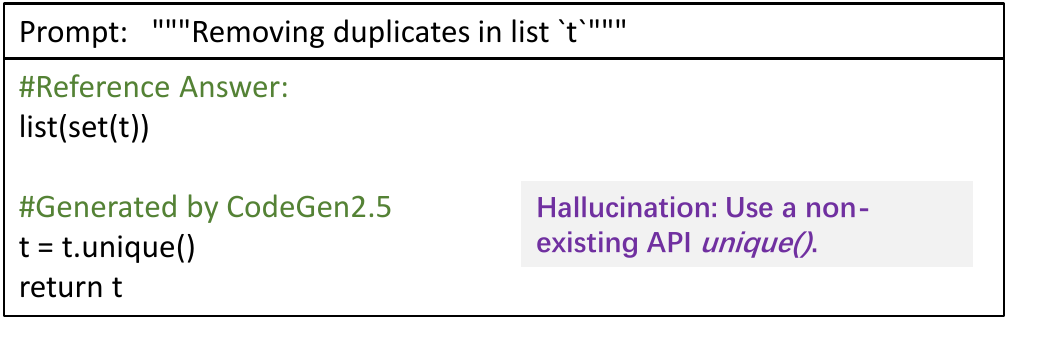}
    \caption{An example of LLM Generating Code with Non-existent API Calls.}
    \label{fig:APIPrompt}

\end{figure}

The issue of API recommendation has garnered significant attention, with several notable studies contributing to the field \cite{9402592, 9793955, 9329198}. However, as illustrated by the simple example in Fig.~\ref{fig:APIPrompt}, CodeGen2.5's model erroneously invokes a non-existent API, \emph{unique()}. Such errors, which can be described as hallucinations, are typically easy to identify using a compiler geared towards programming languages. 

Specially, we observed that PLMs struggle to generate high-quality code when prompts specify the usage of third-party libraries, such as \emph{Pandas}, \emph{NumPy}, \emph{matplotlib}, \emph{Django}, \emph{SQL} and so on. In the 1,264 problematic cases, there are 671 ones involving third-party library usage errors (about 53.09\%). On the other hand, we can see that the introduction of more rigorous benchmarks, such as BigCodeBench \cite{zhuo2024bigcodebench}, is poised to enhance the evaluation of LLMs in code generation tasks that involve a variety of function calls. This, in turn, is expected to propel advancements in LLMs' ability to generate code with accurate API usage even towards complex instructions.

\noindent \textbf{Type VI. Lack of domain knowledge. } 
Due to a lack of common knowledge in certain areas (e.g., IP address ranges, URL format or the definition of radial ticks), the generated code may contain errors or incomplete part. 

\begin{figure}[!htbp]
    \centering
    \includegraphics[trim={0 1cm 16cm 0},clip, width = 0.5\textwidth]{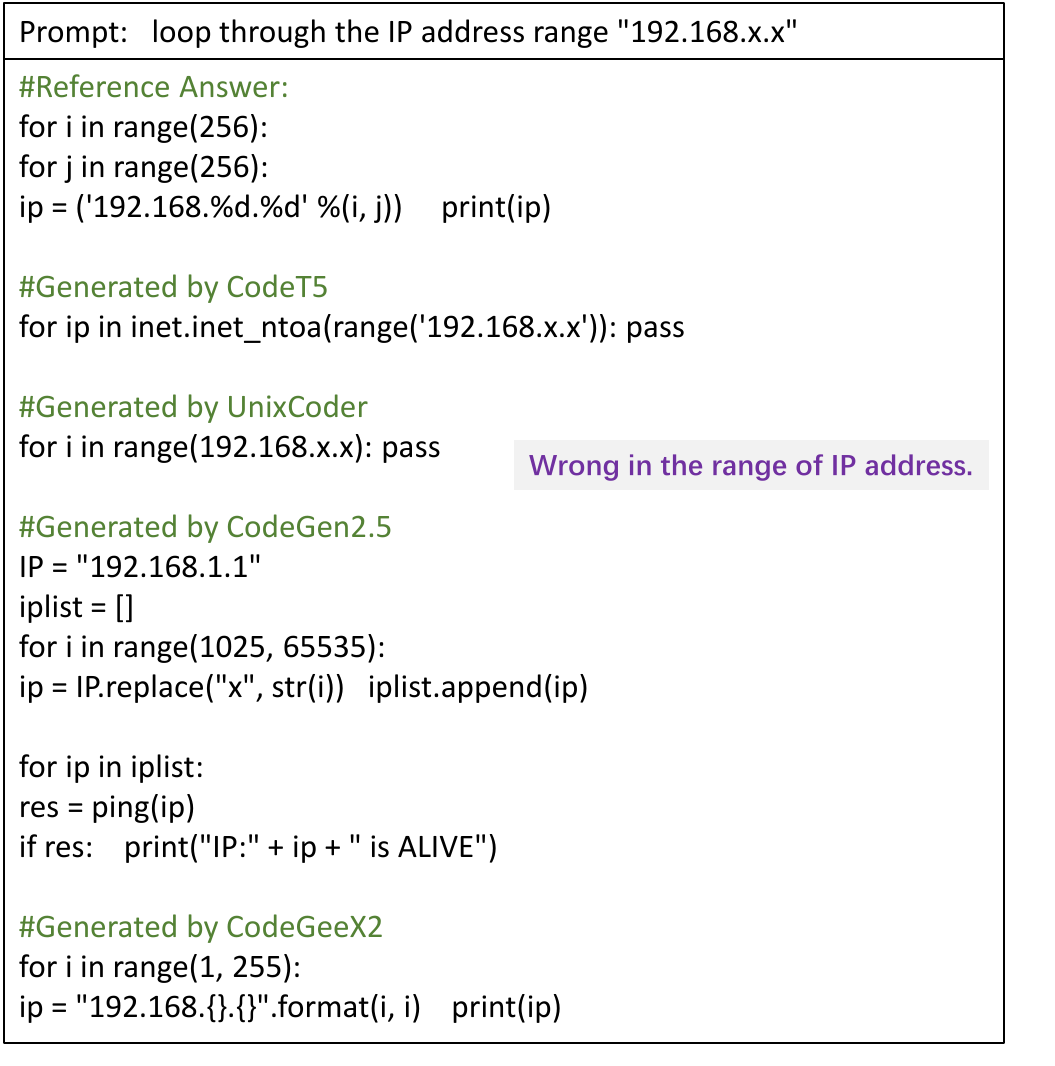}
    \caption{An Example Demonstrating Potential Domain Knowledge Deficiency in LLM during Code Generation.}
    \label{fig:domainKnowledgeExp}
\end{figure}

For instance, PLMs may not inherently understand that IP addresses consist of four numeric segments, each ranging from 0 to 255. Without this knowledge, they could err when generating source code intended to handle IP address ranges. Our observations, drawn from the example in Fig.\ref{fig:domainKnowledgeExp}, indicate that CodeT5, UnixCoder, CodeGen2.5, and CodeGeex2 exhibit shortcomings in generating source code that accurately represents IP address ranges. On the other hand, cutting-edge models like GPT-4 might encapsulate this type of ubiquitous and elementary knowledge, thereby enhancing the precision of generated code for tasks involving IP addresses.

\noindent \textbf{Type VIII. Gold plating. } 
This particular shortcoming, prevalent among LLMs like CodeGen2.5, CodeGeeX2 and ChatGPT, has been highlighted by our research. We use the term ``gold plating'' metaphorically to describe instances where an LLM produces code that is more elaborate or intricate than necessary to meet the specified prompt. During code generation tasks, these models are prone to introducing superfluous features or enhancements—elements that were never solicited by the prompts. 

This issue is well-recognized within the field, since numerous studies adopting truncation strategies for refining the generated output \cite{chen2021evaluating, nijkamp2023codegen2}. Popular methods prioritize the initial segments of generated code, truncating at predetermined stop sequences like `$\backslash nclass$', `$\backslash ndef$', and `$\backslash nprint$. Such approaches, rather than preventing the generation of unnecessary functions or statements, act as a filtering process to remove them post-generation \cite{chen2021evaluating}. 
While these truncation strategies are adept at removing a substantial portion of the unnecessary code, they are often too generalized to discern all vital components with precision. Their \emph{one-size-fits-all} nature can lead to the inadvertent exclusion of nuanced, context-specific elements that are essential for the code's functional integrity. Furthermore, such crude filtering may not account for the logical flow and dependencies within the code, potentially disrupting the cohesive structure necessary for the code to operate as intended.

\begin{figure}[!htbp]
    \centering
    \includegraphics[trim={0 7cm 13cm 0},clip, width = 0.5\textwidth]{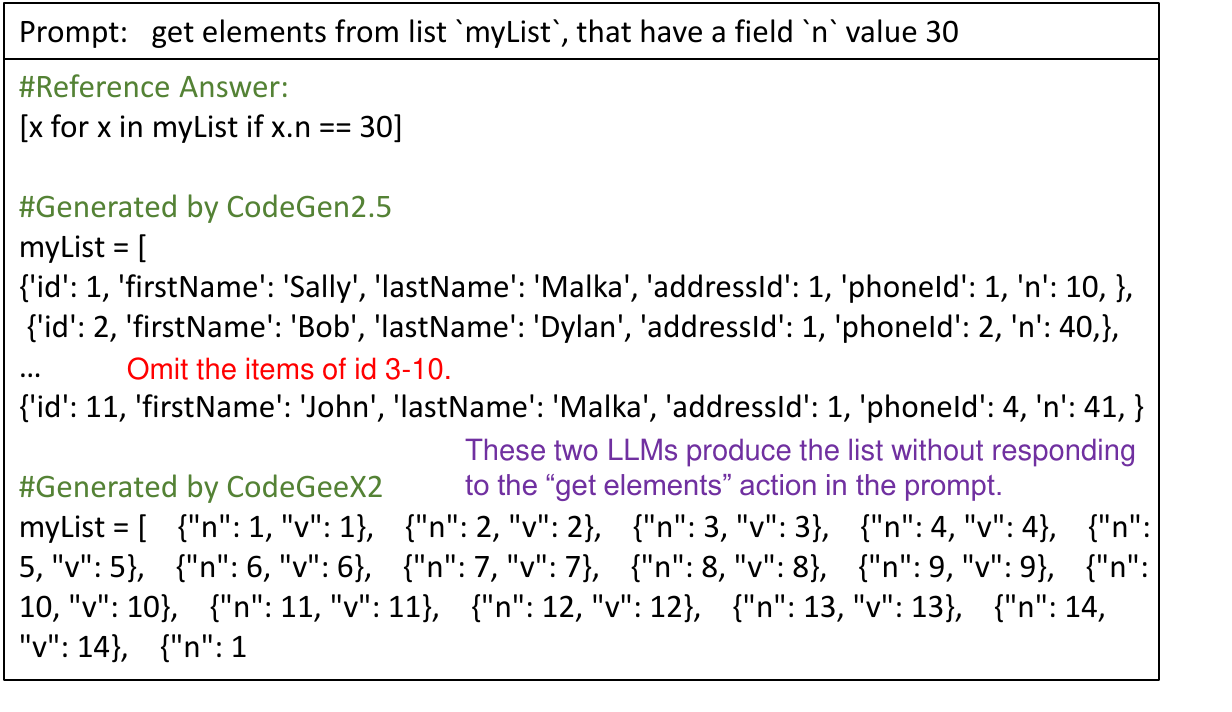}
    \caption{A Demonstrative Example of Feature Overgeneration by LLMs.}
    \label{fig:goldPlating}
\end{figure}

In our study, we employed the truncation methodology delineated in Nijkamp et al. \cite{nijkamp2023codegen2}, which offers a more robust analytical framework than that of Chen et al. \cite{chen2021evaluating}. We observed several instances where the generated code failed to incorporate the core operation specified in the prompt. 
For instance, in the example in Fig.\ref{fig:goldPlating}, both CodeGen2.5 and CodeGeeX2 were found to populate the list `myList' without accurately performing the `get elements' action as stipulated by the prompt. In contrast, GPT-4 predominantly adheres to the prompt's specifications, albeit it additionally assigns the retrieved elements to a pre-defined variable `selected\_elements'.

\noindent \textbf{Type IX. Code duplication. } This represents another shortcoming that we've identified in LLMs, specifically within the scope of our evaluation of CodeGen2.5 and CodeGeeX2. These models occasionally produce redundant code in response to certain prompts, where the repetition may or may not be pertinent to the given task. 

\begin{figure}[!htbp]
    \centering
    \includegraphics[trim={0 6cm 21cm 0},clip, width = 0.35\textwidth]{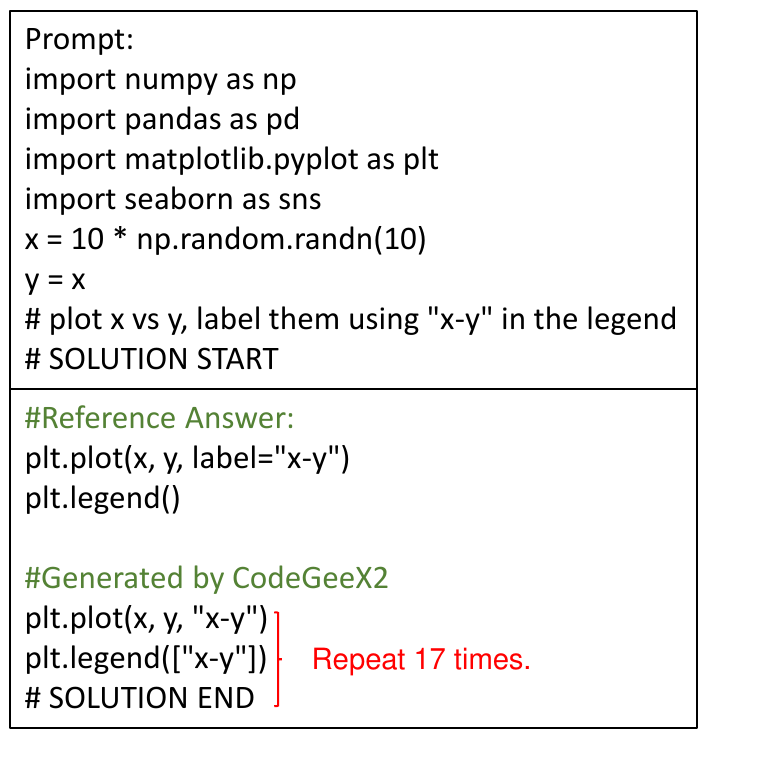}
    \caption{Example of One Model Producing Repeated Code at Output Limit.}
    \label{fig:repeatedGeneration}
\end{figure}

In the example in Fig.\ref{fig:repeatedGeneration}, CodeGeeX2 comes close to generating accurate source code with exceptions, like an improper assignment to the \emph{labels} parameter within the plt.plot() function. The model has a tendency to output nearly identical solutions, repeating them up to 18 times until reaching the maximum output length defined by default. In some scenarios, the repetition of nearly-correct solutions could be considered tolerable since they may still offer valuable guidance for programming assistance. However, this propensity can also become annoying, especially when models generate an excessive amount of redundant and irrelevant source code. In these instances, the repeated inclusion of marginally relevant code acts as nothing more than noise, detracting from the user's experience by cluttering the output with unnecessary information.
   

%% file: sections/Distributions.tex
\section{Weakness distribution over datasets}
We examine the distribution of weaknesses across three benchmarks for the PLMs—three large-scale models and two smaller ones, respectively. Given the remarkable performance of the larger models, we place greater emphasis on their analysis.

By assessing the distribution of weaknesses, our aim is to uncover the predominant types of weaknesses and their proportions within each benchmark. This approach seeks to lay the groundwork for targeted research efforts. For example, research aimed at generating source code from complex instructions could concentrate on benchmarks with a higher incidence of such prompts.

\subsection{Distribution on the bigger models.} 
In this section, we would like to show the distribution over each individual model, and the \emph{overview} distributions through dual lenses: the \emph{union} and \emph{intersection} perspectives.


\subsubsection{Individual model view}
 \label{subsubsec:individual}

Table \ref{tab:singleBigModelDistribution} delineates the proportion of each weakness type for each model across the three benchmarks.
Firstly, GPT-4 distinctly outperforms the other models when it comes to the count of problematic cases, exhibiting fewer issues in both HumanEval+ and DS-1000 than CodeGeeX2 and CodeGen2.5. Within this comparison, CodeGeeX2 shows superior performance relative to CodeGen2.5, which corroborates the results documented in Table \ref{tab:results}.

Secondly, while Type IV weaknesses—primarily misjudgments—predominate for GPT-4 in CoNaLa, this model otherwise demonstrates smaller proportions of all other weakness types across the benchmarks, signaling its overall strength.

Type V, indicative of a failure to completely capture the semantics of the prompts, is the most common weakness identified across all three benchmarks for the models, except for GPT-4 in CoNaLa where it ranks third. This reveals how vital natural language understanding (NLU) is in code generation and suggests that enhancing NLU is an ongoing challenge for current LLMs. Notably, GPT-4 exhibits fewer Type V weaknesses than its peers, reflecting its more advanced NLU proficiency.

It's interesting to observe that all models present multiple instances of Type III `gold plating' within CoNaLa and DS-1000. Yet, no such cases were found in HumanEval+, the only manual curated benchmark in our selected set. This implies that clearer prompt descriptions may assist LLMs in setting appropriate boundaries for the code generation process.

CodeGeeX2 appears particularly susceptible to producing code duplication, displaying this issue consistently across all three benchmarks, with a notable prevalence in CoNaLa and HumanEval+. CodeGen2.5, on the other hand, is most associated with code duplication in DS-1000. In sharp contrast, GPT-4 avoids generating repetitive, redundant source code.

  \begin{table*}[!htbp]
        \centering
        \caption{Distribution of Weakness Types Across Each of the Three Large-Scale Models (\%).}
        \footnotesize
        \label{tab:singleBigModelDistribution}
      \begin{tabular}{|c|c|c|c|c|c|c|c|c|c|c|}
        \hline
        \multirow{2}{*}{\textbf{Model}} & \multirow{2}{*}{\textbf{Benchmark (\#Prob.Cases)}} & \multicolumn{9}{c|}{\textbf{Weakness Types}} \\ \cline{3-11}
        && I&II&III&IV&V&VI&VII&VIII&IX \\ \hline
        \multirow{3}{*}{GPT-4} &\cellcolor{blue!15}CoNaLa (339) & \cellcolor{blue!15}27.14&\cellcolor{blue!15}0.88&\cellcolor{blue!15}0.00&\cellcolor{blue!15}\textbf{43.66} &\cellcolor{blue!15}8.85 &\cellcolor{blue!15}	0.59 &\cellcolor{blue!15}	1.18 &\cellcolor{blue!15}5.60 &	\cellcolor{blue!15}0.00\\ \cline{2-11}
        
        &\cellcolor{green!15}HumanEval+ (38) &\cellcolor{green!15}0.00 &\cellcolor{green!15} 0.00 & \cellcolor{green!15}0.00 &\cellcolor{green!15}0.00 & \cellcolor{green!15}\textbf{42.11} &\cellcolor{green!15}0.00 &\cellcolor{green!15}2.63&\cellcolor{green!15}0.00&\cellcolor{green!15}0.00  \\ \cline{2-11}
        
        &\cellcolor{red!10}DS-1000 (189) &\cellcolor{red!10}8.47&\cellcolor{red!10}7.41&\cellcolor{red!10}0.00&\cellcolor{red!10}0.00&\cellcolor{red!10}\textbf{55.56}&\cellcolor{red!10}1.59&\cellcolor{red!10}0.00&\cellcolor{red!10}1.59&\cellcolor{red!10}0.00 \\ \hline

        \multirow{3}{*}{CodeGeeX2} &\cellcolor{blue!15}CoNaLa (339) & \cellcolor{blue!15}27.14&\cellcolor{blue!15}1.18&\cellcolor{blue!15}	0.00&\cellcolor{blue!15}	32.15&\cellcolor{blue!15}\textbf{39.82}&\cellcolor{blue!15}	2.36&\cellcolor{blue!15}	2.65&\cellcolor{blue!15}	10.62&\cellcolor{blue!15}4.13 \\ \cline{2-11}
        &\cellcolor{green!15}HumanEval+ (82) &\cellcolor{green!15}0.00&\cellcolor{green!15}0.00&\cellcolor{green!15}0.00&\cellcolor{green!15}0.00&\cellcolor{green!15}\textbf{73.71}&\cellcolor{green!15}0.00&\cellcolor{green!15}3.66&\cellcolor{green!15}0.00&\cellcolor{green!15}1.22 \\ \cline{2-11}
        
        &\cellcolor{red!10}DS-1000 (215) &\cellcolor{red!10}6.51&\cellcolor{red!10} 7.91&\cellcolor{red!10}0.00&\cellcolor{red!10}0.00&\cellcolor{red!10}	\textbf{77.21} &\cellcolor{red!10}	6.51 &\cellcolor{red!10}	0.00&\cellcolor{red!10}	2.33&\cellcolor{red!10}	0.47 \\ \hline

        \multirow{3}{*}{CodeGen2.5} &\cellcolor{blue!15}CoNaLa (339) &\cellcolor{blue!15}27.43&\cellcolor{blue!15} 0.88&\cellcolor{blue!15}0.00&\cellcolor{blue!15}	23.01&\cellcolor{blue!15}	\textbf{54.28}&\cellcolor{blue!15}	2.36&\cellcolor{blue!15}	2.06&\cellcolor{blue!15}	8.55&\cellcolor{blue!15}	0.00 \\ \cline{2-11}
        
        &\cellcolor{green!15}HumanEval+ (91) &\cellcolor{green!15}0.00 &\cellcolor{green!15}0.00&\cellcolor{green!15}	0.00&\cellcolor{green!15}0.00 &	\cellcolor{green!15}\textbf{68.13}&\cellcolor{green!15}0.00&	\cellcolor{green!15}2.20&	\cellcolor{green!15}0.00&\cellcolor{green!15}	0.00 \\ \cline{2-11}
        
        &\cellcolor{red!10}DS-1000 (261) &\cellcolor{red!10}6.90&\cellcolor{red!10}6.51&\cellcolor{red!10}0.00&\cellcolor{red!10}0.00&\cellcolor{red!10}	\textbf{75.48}&\cellcolor{red!10}	2.30&\cellcolor{red!10}	0.00&\cellcolor{red!10}	4.21&\cellcolor{red!10}	0.77 \\ \hline

\end{tabular}
\end{table*}

\subsubsection{Overview distribution: union and intersection}
\label{subsubsec:overview}
\textbf{Union distribution.}
In the context of the \emph{union} view, we conceptualize the three large models as a singular, overarching virtual model. For each prompt-oriented record (i.e., the set of prompt-code pairs towards the same prompt), we meticulously examine the presence of weakness types across all three models. Any weakness type identified in at least one model is recorded on the overview weakness table, ensuring that no potential vulnerability is overlooked. 

This union perspective offers a conservative yet exhaustive representation of the weakness distribution, akin to casting the widest possible net to capture every instance of vulnerability. It provides a broad spectrum analysis that anticipates the fullest range of weaknesses that could emerge from these models. As such, it serves as an indispensable comprehensive guide, directing the meticulous scrutiny of source code produced by LLMs. By aggregating the insights gleaned from all models, this approach equips us with a robust framework for preemptively identifying and addressing points of failure, fortifying the reliability of source code generated by LLMs.

Table \ref{tab:bigModelDistributionUnion} shows the \emph{Union} distribution results over the three benchmarks. For better illustration, we highlight the weakness types with ratio of above 50\% with light gray, and those more than 10\% and less than 50\% with light blue.
\begin{table*}[htbp]
        \centering
        \caption{Distribution of Weakness Types Across the Three Large-Scale Models on Three Benchmarks (\emph{Union}).}
        \footnotesize
        \label{tab:bigModelDistributionUnion}
      \begin{tabular}{|p{1.4cm}|p{1.8cm}|p{1.0cm}|p{1.2cm}|p{0.9cm}|p{1.0cm}|p{1.1cm}|p{1.1cm}|p{1.0cm}|p{1.1cm}|}
        \hline
        \textbf{Benchmark} &\multicolumn{4}{c|}{\textbf{Benchmark-based weaknesses}} & \multicolumn{5}{c|}{\textbf{Model-related weaknesses}} \\ \hline
        & \makecell[l]{I. Inaccurate desc.\\(I.i,I.ii, I.iii)} &  \makecell[l]{II. Overly\\ complex\\ prompts} & \makecell[l]{III. Grammar \\errors} & \makecell[l]{IV. Single \\answer\\ biases} & \makecell[l]{V. Missing\\ pivotal\\ semantics} & \makecell[l]{VI. Wrong \\API usage} & \makecell[l]{VII. Lack \\of domain \\knowledge} & \makecell[l]{VIII.Gold\\ plating} &\makecell[l]{IX. Code \\duplication} \\ \hline
        CoNaLa (339) &\cellcolor{blue!15} \makecell[l]{27.43\% (6.78\%, \\17.11\%, 6.19\%)}	&1.18\%	&0.00\%	& \cellcolor{gray!30} 53.10\%	&\cellcolor{gray!30} 65.78\% &	4.72\% &3.24\%	&\cellcolor{blue!15} 13.27\%	&4.13\% \\ \hline
        HumanEval+ (115) & 0.00\%	&0.00\%	&0.00\%	&0.00\%	&\cellcolor{gray!30} 66.09\%	&0.00\%	&5.22\% &	0.00\%	&0.87\% \\ \hline
        DS-1000 (277) & \makecell[l]{6.72\% (0.00\%, \\0.75\%, 5.97\%)} &6.14\%	&0.00\%& 0.00\%	&\cellcolor{gray!30} 80.51\%	&6.86\%&	0.00\%	&5.42\%	&1.08\%  \\ \hline
        
\end{tabular}
\end{table*}

\textbf{Intersection distribution.}
For \emph{intersection} distribution view, we focus on revealing shared weaknesses across the three examined LLMs. By identifying these common vulnerabilities, we highlight areas of critical weakness which demand immediate attention. This approach not only underscores the most pressing weakness concerns but also paves the way for a more targeted and efficient source code review process in practical applications. 

In constructing this intersection distribution, we tag a weakness type only when all three models consistently exhibit that specific vulnerability. The implications of this strategy are significant; it allows us to concentrate our resources on mitigating the most prevalent risks, thereby strengthening the overall robustness of LLMs against potential exploits.

Table \ref{tab:bigModelDistribution-intersection} illustrates the distribution of nine distinct weakness types across three datasets. To enhance visual clarity, we employ a color-coding system: the most salient weakness types are shaded dark gray, the secondary ones in light blue, and the tertiary with a light green background. Upon reviewing the ratios, these three categories fall within the ranges of [20\%, $\infty$), [10\%, 20\%), and [4\%, 10\%), respectively.

\begin{table*}[htbp]
        \centering
        \caption{Distribution of Weakness Types Across the Three Large-Scale Models on Three Benchmarks (\emph{Intersection}).} 
        \scriptsize
        \footnotesize
        \label{tab:bigModelDistribution-intersection}
      \begin{tabular}{|p{1.4cm}|p{1.6cm}|p{1.0cm}|p{1.2cm}|p{1.0cm}|p{1.1cm}|p{1.1cm}|p{1.1cm}|p{1.0cm}|p{1.1cm}|}
        \hline
        \textbf{Benchmark} &\multicolumn{4}{c|}{\textbf{Benchmark-based weaknesses}} & \multicolumn{5}{c|}{\textbf{Model-related weaknesses}} \\ \hline
        & \makecell[l]{I. Inaccurate desc.\\(I.i,I.ii, I.iii)} &  \makecell[l]{II. Overly\\ complex\\ prompts} & \makecell[l]{III. Grammar \\errors} & \makecell[l]{IV. Single \\answer\\ biases} & \makecell[l]{V. Missing\\ pivotal\\ semantics} & \makecell[l]{VI. Wrong \\API usage} & \makecell[l]{VII. Lack \\of domain \\knowledge} & \makecell[l]{VIII.Gold\\ plating} &\makecell[l]{IX. Code \\duplication} \\ \hline
        CoNaLa (339) & \cellcolor{gray!70} \makecell[l]{26.84\% (6.49\%, \\16.52\%, 6.19\%)} & 0.88\% & 0.00\% & \cellcolor{blue!15} 
 12.09\% & \cellcolor{green!10} 5.60\% & 0.00\% & 0.59\% & \cellcolor{green!10} 5.60\% & 0\%\\ \hline
        HumanEval+ (115) & 0.00\% & 0.00\% &	0.00\% &	0.00\% 	&\cellcolor{blue!15} 11.30\% &0.00\% &	0.00\% &0.00\% &	0.00\% \\ \hline
        DS-1000 (277) & \cellcolor{green!10} \makecell[l]{4.33\% (0.00\%, \\0.00\%, 4.33\%)} & \cellcolor{green!10} 5.05\% &0.00\% & 0.00\% &	\cellcolor{gray!70} 24.91\% &	0.00\%&	0.00\% &0.36\% & 0\% \\ \hline
\end{tabular}
\end{table*}

\textbf{Observations and discussion.}
Looking at the results in Table \ref{tab:bigModelDistributionUnion} and Table \ref{tab:bigModelDistribution-intersection}, we have the following four observations.
\begin{itemize}[leftmargin = 0.4cm, itemindent = 0.1cm]
  \item  \emph{Type V: Missing pivotal semantics}, is the most prevalent weakness among the three larger models, whether we consider the intersection or union perspective. Specifically, Table \ref{tab:bigModelDistributionUnion} shows that at least one model fails to include some crucial semantic elements in 65.78\% of all CoNaLa samples, 66.09\% in HumanEval+, and 80.51\% in DS-1000. Similarly, Table \ref{tab:bigModelDistribution-intersection} reveals that each of the three models exhibits this type of shortcoming in precisely 11.30\% of HumanEval+ samples and 24.91\% of DS-1000 samples.

Towards the three benchmark, this issue is most pronounced in DS-1000, possibly due to the complexity of its prompts. The prompts in DS-1000 are not only the longest, as shown in Table \ref{tab:dataset}, but each of them also incorporate an ``event flow''—the original intent, failed attempts with outcomes, and the current status. While this provides a richer context, it may introduce additional noise that complicates the models' understanding of the prompts' true intentions.

\item \emph{Type I: Inaccurate Description} is another issue that can not be ignored. It stands out as a significant weakness in CoNaLa. Both referenced tables indicate that all three models produce erroneous codes for the same 26.84\% of samples, which have been identified as inaccurately described.

Compared to the other two datasets, CoNaLa's prompts are the briefest in terms of token count (refer to Table \ref{tab:dataset}), suggesting simpler tasks. However, this brevity may result in a lack of explanatory detail and context, impeding the models' ability to interpret ambiguous or incomplete semantics in CoNaLa prompts.

In addition to CoNaLa, DS-1000 also exhibits this type of weakness. Problematic code is produced by at least one model for 6.72\% of the prompts, all of which have been identified as inaccurately described. More notably, 4.33\% of all samples are consistently problematic across all three models due to their inaccuracy.


\item \emph{Type VIII: Gold Plating} also warrants significant attention, particularly in CoNaLa and DS-1000, with the problem being more severe in CoNaLa.

This trend may stem from a common underlying factor: the relatively simplistic nature of the problems presented, in contrast to the more complex ones found in HumanEval+. LLMs appear prone to over-elaboration when faced with uncertain responses. This is supported by the presence of 5.42\% Type VIII weaknesses in DS-1000, which has shorter average token lengths compared to HumanEval+. While the absence of Type VIII weaknesses in HumanEval+ could be linked to more complex variables—such as superior quality of prompt descriptions (i.e., no Type I issues)—it is noteworthy that responses in HumanEval+ tend to be more verbose.

\item \emph{Type IV: Single Answer Biases} As a notable benchmark lacking test cases, it's unsurprising that Type IV: Single answer biases is prominently observed in CoNaLa. In particular, at least one model generates ``problematic code'' for 53.10\% of all prompts, yet these responses are deemed correct upon manual inspection.

\end{itemize}

\subsection{Distribution over two smaller models (Intersection)}

We have also analyzed the intersection of weakness distributions for the two smaller models, with results presented in Table \ref{tab:smallIntersection}. For each dataset, the most frequent weaknesses are highlighted with a light gray background, and the second most frequent are denoted with a light blue background.

A comparison with the findings in Table \ref{tab:bigModelDistribution-intersection} on the large models reveals an overall increase in the distribution of most types of weaknesses for the smaller models, with the exception of Type I in the DS-1000 dataset.

Despite the variance in sample sizes used for analyzing large and small models, the different distributions observed suggest a more pronounced prevalence of weaknesses in the code generated by smaller models. For instance, smaller models seem to be more sensitive to the precision of prompt descriptions (Type I), as evidenced by the difference in percentages—40\% versus 26.84\% in CoNaLa and 6.25\% versus 0 in HumanEval+. While Type I was not detected in DS-1000, this does not imply that the quality of prompt descriptions is less influential in this dataset. There are two possible explanations: firstly, the fine-tuning process on smaller models might have enhanced their adaptability to the dataset; secondly, the fine-tuning resulted in fewer samples for evaluating smaller models, meaning we can only categorize the weaknesses present in these limited samples. It is conceivable that samples with Type I.iii (inconsistency) were excluded from the analysis. Moreover, smaller models face greater challenges in providing accurate answers to overly complex prompts—a finding that aligns with our intuition. In benchmarks featuring longer prompts, the occurrence of overly complex issues increases significantly, rising from 0 to 43.75\% in HumanEval+ and from 5.05\% to 27.00\% in DS-1000.

In terms of model-related weaknesses, there is a higher incidence of generated code missing pivotal semantics in the prompts (Type V), with this issue being more severe in benchmarks characterized by longer prompts. Additionally, a larger proportion of samples exhibit Type VI weaknesses, indicating that even when the generated code semantically matches the corresponding prompts, there may still be incorrect API usage or the invocation of non-existent APIs.

 \begin{table*}[htbp]
        \centering
        \caption{Distribution of Weakness Types Across the Two \emph{Smaller Models} on Three Benchmarks (\emph{Intersection})}.
        \footnotesize
        \label{tab:smallIntersection}
      \begin{tabular}{|p{2.2cm}|p{2.0cm}|p{1.1cm}|p{1.3cm}|p{1.2cm}|p{1.3cm}|p{1.5cm}|p{1.6cm}|}
        \hline
    \textbf{Benchmark} &\multicolumn{4}{c|}{\textbf{Benchmark-based weaknesses}} & \multicolumn{3}{c|}{\textbf{Model-related weaknesses}} \\ \hline
        & \makecell[l]{I. Inaccurate desc.:\\I.i vague.\\I.ii incomplete. \\I.iii inc.} &  \makecell[l]{II. Overly\\ complex\\ prompts} & \makecell[l]{III. Grammar \\errors} & \makecell[l]{IV. Single \\answer\\ biases} & \makecell[l]{V. Missing\\ pivotal\\ semantics} & \makecell[l]{VI. Wrong \\API usage} & \makecell[l]{VII. Lack \\of domain \\knowledge} \\ \hline
        CoNaLa (500) & \cellcolor{gray!70} \makecell[l]{40.00\% (13.00\%,\\24.40\%, 5.40\%)} & 3.00\%  & 1.00\% & \cellcolor{gray!70} 40.80\% &  14.80\% & \cellcolor{blue!15}38.40\%	& 3.00\% \\ \hline
        HumanEval+ (16) &  \makecell[l]{6.25\% (6.25\%, \\0.00\%, 0.00\%)} & \cellcolor{blue!15} 43.75\%  & 0.00\% & 0.00\%\% & \cellcolor{gray!70} 75.00\% &  6.25\%	&  6.25\%\\ \hline
        DS-1000 (100) &  0.00\% & \cellcolor{blue!15} 27.00\%  & 0.00\% & 0.00\% & \cellcolor{gray!70} 51.00\% & \cellcolor{gray!70}69.00\%	&  0.00\%\\ \hline

\end{tabular}
\end{table*}

\section{Manually Curating Prompts of CoNaLa}
\label{sec:CoNaLaCurating}

As discussed in Section \ref{subsubsec:overview}, within the CoNaLa dataset, each of the three larger models encounters difficulties generating correct source code for the same subset of prompts—specifically, 26.84\%—exhibiting Type I weaknesses (as detailed in Table \ref{tab:bigModelDistribution-intersection}). Additionally, a higher incidence of this weakness, approximately 40\%, is noticeable in the code produced by the two smaller models (as shown in Table \ref{tab:smallIntersection}).

To investigate the effects of enhanced prompt quality on the generated source code, we aim to refine the prompt design by addressing these identified weaknesses. To this end, we have curated a set of 90 problematic prompts. We will manually revise both the prompts and their reference answers to mitigate the weaknesses and then evaluate the performance of the five models using both the original and improved prompts.

For Type I.i and Type I.ii, we modify the prompts to align with user intentions while keeping changes to the original text minimal. For Type I.iii, the reference code will be adjusted if it is found to conflict with the prompt. The guiding principle for modifying prompts is to ensure they continue to embody the user's intention without delving into implementation details. 

\begin{figure}[!htbp]
    \centering
    \includegraphics[trim={0 11cm 19cm 0},clip, width = 0.4\textwidth]{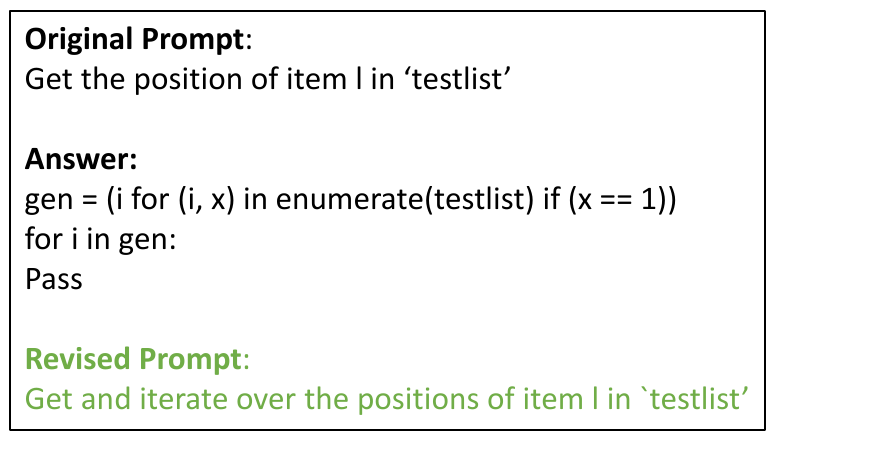}
    \caption{Example Showing the Result of Manual Curating of One Prompt.}
    \label{fig:promptCurating}
\end{figure}

As an illustration, in the case of the Fig.\ref{fig:promptCurating}, the prompt should not incorporate the iteration segment, `for i in gen:  pass', that appears in the answer. To enhance the clarity and completeness of the prompt while preserving its intent, we suggest revising it to \emph{`get \textcolor{blue}{and iterate over} the positions of item 1 in `testlist'}. This refinement steers clear of rewriting the prompt to an overly technical description such as \emph{`Iterate through `testlist' and create a generator to yield all indices where the item is exactly 1, then exhaust this generator without performing any operations on the retrieved indices.'}

Each of the first three authors were assigned 30 samples, each with the prompt description, the reference source code and the identified weakness types. Once the modifications have been finished, they performed a mutual review, i.e., the second author reviewed the result of the first author, the third author reviewed that of the second author, and the first author reviewed that of the third author. The reviewer checked that whether the identified weaknesses are identified, and whether unnecessary code information was contained in the prompts. They marked the discrepancy, and the three authors would discuss jointly until an agreement has been achieved. 

We subsequently re-evaluated the three larger models using the updated prompts. The smaller models were excluded from this round of testing because some of these 90 records had been utilized in their fine-tuning process. Table \ref{tab:promptRevisionCompare} presents the Exact Match (EM) and CodeBLEU scores for both the original and revised prompts, clearly demonstrating enhanced performance across all three models on the refined benchmark. For a clearer depiction, we calculated the gains in EM and CodeBLEU  and highlighted them in red. Notably, GPT-4 shows the greatest sensitivity to the prompt modifications, with gains of 400\% in EM and 29\% in CodeBLEU.  While the EM for the code generated by CodeGen2.5 and CodeGeeX2 on the polished prompts remains unchanged—suggesting that the proportion of codes exactly matching the reference answer has not been affected—the CodeBLEU score improvements are nevertheless substantial, with gains of 12\% for CodeGen2.5 and 7\% for CodeGeeX2, respectively.

 \begin{table}[htbp]
        \centering
        \caption{Comparisons of the Model Performance on the Original and Revised prompts.}.
        \renewcommand \arraystretch{1.5}
        \footnotesize
        \label{tab:promptRevisionCompare}
      \begin{tabular}{|c|c|c|c|c|}  \hline
      \multirow{2}{*}{Models} & \multicolumn{2}{c|}{Original Prompts} & \multicolumn{2}{c|}{Revised Prompts} \\ \cline{2-5}
      &EM & CodeBLEU &EM & CodeBLEU \\  \hline
      CodeGen2.5 &0.004 & 0.25 & 0.004 (-) &0.28 (\textcolor{red}{$\uparrow12\%$}) \\ \hline
      CodeGeeX2 & 0.00 & 0.26 & 0.00 (-) & 0.28 (\textcolor{red}{$\uparrow7\%$})\\ \hline
      GPT-4 & 0.01 & 0.31 & 0.05 (\textcolor{red}{$\uparrow400\%$}) &0.40 (\textcolor{red}{$\uparrow29\%$})\\ \hline
\end{tabular}
\end{table}

%% file: sections/Discussion.tex
\section{Discussion}
\label{sec:discussion}

\subsection{Limitations and Opportunities} 
\label{subsec:suggestions}

\begin{itemize}[leftmargin = 0.4cm, itemindent = 0.1cm]
    \item \noindent \textbf{Paradigm of Effective Prompting for Code-PLMs is unclear.} It is widely acknowledged that well-crafted prompts are essential for generating high-quality code. Yet, defining what constitutes an effective prompt for code-centric code-PLMs remains elusive. Based on our observations (see Table \ref{tab:singleBigModelDistribution}), inaccurate prompt description ranks among the top three types of weaknesses across all benchmarks and for each of the three big PLMs studied. Despite the existence of some recent work, the exploration of prompt patterns has predominantly been from a human perspective \cite{white2023prompt, mishra-etal-2022-reframing}. These studies provide intriguing insights but are still preliminary; systematic evaluations across more diverse benchmarks and code-PLMs are imperative prior to their widespread adoption.

Moreover, there is an increasing trend in research aimed at improving existing prompts, employing techniques such as prompt clarification \cite{li2023python} and prompt decomposition \cite{khot2023decomposed, dua2022successive}. Yet, a definitive, `silver bullet' approach remains elusive. The ongoing quest to perfect prompt engineering continues to inspire and challenge both researchers and practitioners in the field.

\item \noindent \textbf{Fine-grained representation of pivot semantics in prompts lacks attention.} 
While numerous code-generation methodologies have been directed towards modeling source code from various angles \cite{sun2019grammar,Brockschmidt2019GenerativeCodegen,Zhong2020scaffolds}, there is a paucity of research on accurately capturing the semantic nuances conveyed by prompts.

Our analysis, as detailed in Tables \ref{tab:bigModelDistribution-intersection}, \ref{tab:smallIntersection}, and \ref{tab:singleBigModelDistribution}, indicates that the omission of crucial semantic content in prompts (Type V) is a prevalent issue. This trend persists across all five models evaluated, spanning both smaller and larger PLMs, and is especially evident in the complex benchmarks of HumanEval+ and DS-1000.

Intuition suggests that the significance ascribed to specific terms—particularly those describing core intentions such as actions and their associated constraints—and the experiences of failure should diverge. Therefore, a promising avenue for research lies in distinguishing various semantic components within prompts, with an emphasis on clarifying the core intent. By doing so, we can aspire to construct representations that capture fine-grained semantic details, thereby facilitating the advanced training and prompting of models.

\item \noindent \textbf{Code generation that involves third-party libraries remains a significant challenge.} Despite the advent of promising approaches in previous studies targeting the recommendation of third-party libraries for code generation models \cite{ijcai2022p329}, further exploration is essential in this field. Our detailed analysis of PLMs—as discussed in Section \ref{subsec:modelWeakness}—demonstrates that substantial difficulties persist with these issues. Moreover, the introduction of more rigorous benchmarks such as BigCodeBench \cite{zhuo2024bigcodebench} aims to evaluate the capability of code generation models to navigate diverse function calls and complex instructions. Concurrently, there is increasing research attention toward areas like private library recommendation \cite{zan-etal-2022-language} and unseen API recommendation \cite{zhou2023docprompting}. We are optimistic that the emergence of such benchmarks and focused research efforts will catalyze the creation of more adept models.

\item \noindent \textbf{Enrich benchmark datasets with answers involving multiple diverse APIs.}
It is widely acknowledged that current metrics, whether based on textual/semantic similarity or unit-test execution, are incapable of precisely evaluating code generation performance. Despite the continuous introduction of novel metrics such as CrystalBLEU \cite{eghbali2022crystalbleu} and CodeBERTScore \cite{zhou2023codebertscore}, biases related to single-answer scenarios have yet to receive sufficient attention. Given the prevalence of this issue—for instance, a 43.66\% accuracy rate for GPT-4 across 339 random samples from CoNaLa—we advocate for increased research efforts to establish benchmarks that offer multiple diverse API-based solutions for each prompt, enabling more robust evaluations.
\end{itemize}

\subsection{Threats to Validity}
\label{subsec:threats}

\begin{itemize}[leftmargin = 0.4cm, itemindent = 0.1cm]
    \item \noindent \textbf{Manual annotation.} This procedure might introduce subjective biases. To mitigate this threat, we assembled an annotation team of 14 experienced participants and followed the process of thematic analysis. We conducted a pilot study on small-scale samples to determine an initial weakness category and estimate the annotation cost, with the aim of ensuring most participants' annotation quality. We provided a simple tutorial and a guideline document to the participants before their annotation, ensuring that they understood our task and the initial weakness types.

\item \noindent \textbf{Manual rectification.} To underscore the importance of mitigating the weaknesses in prompts, we conducted a preliminary manual correction process (See Section \ref{sec:CoNaLaCurating}). During this phase, our modifications didn't aim to cater specifically to the standard answer code. Instead, our focus was on promptly following the guidelines and addressing the associated weaknesses as effectively as possible \cite{white2023prompt, mishra-etal-2022-reframing}. The modified prompts have been made publicly accessible via the artifact URL for transparency and further investigation.

\item \noindent \textbf{Implementation correctness of each model.} To ensure the correctness of model implementation, we directly used the source code downloaded from their official websites. Additionally, to optimize the models' performance, we fine-tuned them according to their parameter scales and the similarity between their pre-training datasets and our benchmarks. 

\item \noindent \textbf{Generalization.} Limited by accessibility of PLMs, benchmark datasets, and our research cost, we only analyzed three Python datasets in this initial study. Due to the commonalities among datasets, i.e., prompts, code and the evaluation metrics, our observations should theoretically hold true for other models and programming language-based datasets. However, we have not evaluated the impact of these weaknesses on more models with other programming language datasets.
\end{itemize}

%% file: sections/RelatedWork.tex
\section{Related Work}
\label{sec:relatedWork}

Recent studies have increasingly focused on systematically evaluating popular LLMs for code-related tasks, such as vulnerability detection \cite{steenhoek2022empirical}, program repair \cite{cao2023study}, code summarization \cite{shi2022evaluation}, code completion, and code generation \cite{ciniselli2021empirical, mastropaolo2023robustness, abukhalaf2023codex, zan-etal-2023-large}.

With respect to neural code generation, numerous investigations have scrutinized the bugs in the output. For instance, Liu et al. \cite{10507163} evaluate the code generated by ChatGPT from three aspects: correctness, complexity and security. 
Dou et al. \cite{dou2024whatswrongcodegenerated} examined the performance of seven LLMs over common benchmarks, classifying \emph{bugs} into syntax, runtime, and functional categories based on the Python interpreter's feedback. Wang et al. \cite{wang2024largelanguagemodelsfail} studied incorrect snippets from six LLMs on the HumanEval benchmark, constructing an \emph{error taxonomy} that accounts for issues like missing conditions or incorrect return values. Tambon et al. \cite{tambon2024bugs} identified 10 bug patterns across 333 bugs in code generated by three LLMs, including syntax errors and wrong input types. These studies provide granular insights into the flaws within code produced by LLMs, but tend to \emph{focus solely on the errors in the generated code}. Our current work aims to deliver a more holistic perspective that illuminates weaknesses in code generation, paying close attention \emph{not only to the resultant code but also to the benchmarks utilized—prompt design and reference answers included}.

In addition to examining the bugs, research has explored factors influencing code generation quality. Antonio et al. \cite{mastropaolo2023robustness} performed an empirical study showing that variances in natural language descriptions could affect the output of GitHub Copilot. Cao et al. \cite{cao2022can} observed similar impacts across different datasets and models. Jin et al. \cite{jin2022good} found that carefully crafted prompts significantly improve zero-shot performance, and Lu et al. \cite{lu-etal-2022-fantastically} demonstrated the sensitivity of prompt ordering to few-shot learning success, devising a method to identify effective prompts without additional annotated data.

Shi et al. \cite{shi2022evaluation} emphasized the role of data preprocessing in the performance of state-of-the-art code summarization models. Zan et al. \cite{zan-etal-2023-large} identified three critical factors for LLM success in the NL2Code task: model size, high-quality data, and expert tuning. Liu et al. \cite{liu2023code} pointed out test inadequacy and vague prompt descriptions in benchmarks, proposing EvalPlus, a framework capable of generating a broader set of test cases to expose more nuanced issues. Moradi Dakhel et al. \cite{MORADIDAKHEL2023111734} assessed Copilot against human programmers, noting its shortcomings in synthesizing complex solutions and its occasional neglect of key prompt details—a phenomenon our research categorizes as \emph{Type V: Missing pivotal semantics}.

The aforementioned studies have also provided crucial insights for improving related models. For instance, Seif et al. \cite{abukhalaf2023codex} discovered that UML-enhanced prompts boost the reliability of generated Object Constraint Language (OCL) outputs. Other research has demonstrated the benefits of including context elements, like variable names and function arguments, to refine code generation \cite{Conala}\cite{iyer-etal-2018-mapping}. The impact of external knowledge has been similarly affirmed: Liu et al. \cite{DBLP:conf/acl/0010LLWWBCH22} showed that it enhances commonsense reasoning in sequence models, suggesting that prompting with domain-specific information (Type VII) might improve code generation.

Some researchers have also highlighted benchmark limitations. Liu et al. \cite{liu2024codemind} warned that models potentially overfit on well-trodden datasets like HumanEval, calling for more diverse datasets for authentic assessment of code generation. Huang et al. \cite{huang2024effibench} introduced EffiBench, a new benchmark focusing on the efficiency of model-generated code, urging consideration of not just correctness but also the efficiency and sustainability of code outputs.

%% file: sections/Conclusion.tex
\section{Conclusions and Future work}
\label{sec:conclusions}

To explore the weaknesses in automated code generation, we undertook an empirical study with five SOTA code-PLMs and three benchmark datasets. We conducted a thematic analysis of prompt-code weaknesses, categorizing them into nine types from the comprehensive perspectives of prompt, model, and benchmark quality. We thoroughly examined the distribution patterns of weaknesses in both smaller and larger models through a comprehensive lens that encompasses both intersectional and union perspectives among different models, as well as individual model assessments. We identified a significant occurrence of inaccurate prompts in the CoNaLa dataset and outlined the challenges associated with complex prompts for the two smaller models. Additionally, we highlighted a recurrent issue across all models and benchmarks: the omission of critical semantics necessary for accurate code generation. Moving forward, our research efforts will concentrate on isolating these crucial semantic components within prompts. We intend to refine prompts by incorporating explicit constraints that emphasize these key semantics, anticipating that this focused approach will yield improvements in the quality of the generated source code.